\documentclass[letterpaper, 10pt, journal, twoside]{IEEEtran}
\usepackage{graphics}
\usepackage{stfloats}
\usepackage{epsfig}
\usepackage{times}
\usepackage{amsthm}
\usepackage{amssymb}
\usepackage{amsmath}
\usepackage{indentfirst}
\usepackage{epstopdf}
\usepackage{multirow}
\usepackage{float}
\usepackage{color}
\usepackage{enumerate}
\usepackage{subfigure}
\usepackage{comment}
\usepackage{cite}
\usepackage[ruled,vlined]{algorithm2e}
\usepackage{algpseudocode}
\usepackage{flushend}
\usepackage{amssymb}
\usepackage{stackengine} 
\newcommand\stackrqarrow[2]{%
    \mathrel{\stackunder[0pt]{\stackon[2pt]{$\rightsquigarrow$}{$\scriptscriptstyle#1$}}{%
            $\scriptscriptstyle#2$}}}
\usepackage{pgf}
\usepackage{tikz}
\usetikzlibrary{positioning, arrows.meta,automata}
\usetikzlibrary{calc}
\usepackage[latin1]{inputenc}

\usepackage{tcolorbox}

\definecolor{newGrey}{RGB}{175 238 238}

\title{A Unified Framework for Verification of Observational Properties for Partially-Observed Discrete-Event Systems}

\author{Jianing~Zhao,~\IEEEmembership{Student Member,~IEEE,}
	Xiang~Yin,~\IEEEmembership{Member,~IEEE,}
	Shaoyuan Li,~\IEEEmembership{Senior Member,~IEEE}
	\thanks{This work was supported by the National Natural Science Foundation of China (62061136004, 62173226, 61803259) and by the National Key Research and Development Program of China (2018AAA0101700).
}
\thanks{J. Zhao, X. Yin and S. Li are with the Department of Automation, Shanghai Jiao Tong University, and the Key Laboratory of System Control and Information Processing, the Ministry of Education of China, Shanghai 200240, China. {\tt  E-mail: \{jnzhao,yinxiang,syli\}@sjtu.edu.cn} (Corresponding Author: Xiang Yin)}
}

\newtheorem{mydef}{Definition}

\newtheorem{mythm}{Theorem}

\newtheorem{myexm}{Example}
\theoremstyle{remark}  
\def \trace{\textsf{Trace}}
\def \run{\textsf{Run}}
\def \AP{\mathcal{AP}}
\def \L{\mathcal{L}}
\def \O{\mathcal{O}}
\def \I{\mathfrak{I}}
\def \F{\text{\emph{F}}}
\def \U{\mathcal{U}}
\def \S{\text{\emph{S}}}
\def \N{\text{\emph{N}}}
\def \NS{\text{\emph{NS}}}
\def \NN{\mathbb{N}}
\def \X{\mathbb{X}}

\begin{document}

	\maketitle \pagestyle{plain}
	
\begin{abstract}
In this paper, we investigate property verification problems in partially-observed discrete-event systems (DES). Particularly, we are interested in verifying \emph{observational properties} that are related to the information-flow of the system. Observational properties considered here include  diagnosability, predictability,  detectability and  opacity, which have drawn considerable attentions in the literature. However, in contrast to existing results, where different verification procedures are developed for different properties case-by-case, in this work, we provide a \emph{unified framework} for verifying all these properties by reducing each of them as an instance of \emph{HyperLTL model checking}. Our approach is based on the construction of a Kripke structure that effectively captures the issue of unobservability as well as the finite string semantics in partially-observed DES
so that HyperLTL model checking techniques can be suitably applied. Then for each observational property considered, we explicitly provide the HyperLTL formula to be  checked over the  Kripke structure for the purpose of verification. Our approach is uniform in the sense that all different properties can be verified with the same model checking engine. Furthermore, our unified framework also brings new insights for   classifying  observational properties for partially-observed DES in terms of their verification complexity. 
\end{abstract}
\begin{IEEEkeywords}
Discrete-Event Systems, Partial Observation, Property Verification,  HyperLTL.
	\end{IEEEkeywords}

\section{Introduction}
\subsection{Motivation}
\IEEEPARstart{D}{screte-Event} Systems (DES) is an important class of complex engineering systems with discrete state-spaces and event-triggering dynamics \cite{cassandras2021introduction}. It is widely used in the modeling and analysis of the high-level logic behaviors of complex automated systems such that manufacturing systems, softwares and autonomous robots. Given a DES, one of the most fundamental problems is to determine whether or not the designed system satisfies some desired specifications of our interest by formal and algorithmic procedures. This is also referred to as the \emph{property verification} problem, which is critical to ensure safety and security of DES \cite{liu2022secure}.

In many practical scenarios, DES are usually \emph{partially-observed} either from the system-user's point of view due to the limited sensing capabilities, or from the outsider's point of view due to the partial information release \cite{hadjicostis2020estimation}. In this context, one may need to determine whether or not the observer has sufficient knowledge about the system based on both the DES model and the partial observations. Such properties related to the \emph{information-flow} of the partially-observed DES are referred to as the  \emph{observational properties} \cite{yin2017verification,masopust2019complexity}. 
In this paper, we are concerned with the problems of verifying observational properties for partially-observed DES.

\subsection{Related Works}
Property verification of partially-observed DES dates back to the early investigations of supervisory control of partially-observed DES, where the notion of observability was investigated \cite{lin1988observability,wonham2018supervisory,wonham2019supervisory}. 
In this setting, it is usually assumed that the behaviors of the systems can only be observed partially via a natural projection or an observation mask, and one needs to determine whether or not the imperfect information is sufficient to realize a supervisor.  Later on, verification  of partially-observed DES has been investigated more thoroughly in the contexts of fault diagnosis \cite{lafortune2018history}, fault predication \cite{watanabe2021fault}, state detection \cite{ma2020marking} and security analysis \cite{basilio2021analysis}.  
The reader is referred to the recent textbook \cite{hadjicostis2020estimation} and tutorial paper \cite{yin2019estimation} for more details on this topic. 
In what follows, we briefly review some observational properties that are considered in this work.

One of the most widely investigated observational property in the DES literature is the notion of \emph{diagnosability}, which is initially proposed in \cite{lin1994diagnosability,sampath1995diagnosability} and has been extended to many variants \cite{ran2018codiagnosability,ma2021marking,takai2021general,carvalho2021comparative}.
Specifically, it is assumed that the system is subject to some faults, which are modeled as a set of unobservable events.  Then diagnosability characterizes the ability that one can always determine the occurrences of fault events within a finite  delay. As the dual of the fault diagnosis problem, fault prognosis problem considers to predict the occurrences of faults in advance. In the literature, the notion of prognosability or \emph{predictability} has been adopted   as the necessary and sufficient condition under which fault can always be predicted with no false-alarm or miss-alarm \cite{genc2009predictability,jeron2008predictability,takai2015robust,yin2018verification,chen2021stochastic}. 

In the context of state estimation of partially-observed DES, Shu and Lin proposed several different notions of \emph{detectability} \cite{shu2007detectability,shu2010detectability,shu2012detectability,shu2012delayed} to systematically characterize whether or not the state of the system can be determined unambiguously based on the  information-flow. For example, strong detectability (respectively, I-detectability) requires that, after some finite delays,  the current-state (respectively, initial-state) of the system can always be determined precisely. On the other hand, weak detectability requires that the precise state can be determined along some trajectory of the system. It has been shown that checking strong detectability can be done in polynomial time while checking weak detectability is PSPACE-complete \cite{zhang2017problem,masopust2018complexity,balun2021verification}. 

More recently, motivated by the security and privacy considerations in cyber-physical systems, the notion of \emph{opacity} has drawn many attentions in the literature; see, e.g., \cite{bryans2008opacity,cassez2012synthesis,tong2018current,mohajerani2020compositional,lefebvre2021exposure}.  In this context, it is assumed that there exists a passive intruder (eavesdropper) that can access the information-flow of the system. On the other hand, the system has some ``secret" that does not want to be revealed to the intruder. Then opacity essentially characterizes the plausible deniability of the system such that the intruder can never infer its secret. Depending on different security requirements, different notions of opacity have been studied including, e.g.,  initial-state opacity \cite{saboori2013verification}, current-state opacity \cite{lin2011opacity}, $K$-step opacity \cite{saboori2011verification} and infinite-step opacity \cite{yin2017new,saboori2011verification}. Relationships among different notions of opacity are discussed in  \cite{wu2013comparative,balun2021comparing,wintenberg2022general}.

\subsection{Our Results and Contributions}	
While there is a wide literature on the   verification of observational properties  for partially-observed DES, several problems still remain.  In particular, the existing approaches for the verification of partially-observed DES are mainly based on the observer structure and its variants \cite{cassandras2021introduction}.  For some properties, such as diagnosability, predictability and strong detectability, researchers have further proposed polynomial-time algorithms \cite{jiang2001polynomial,yoo2002polynomial,jeron2008predictability,shu2010detectability}. However, the existing verification techniques are mainly developed for different properties \emph{case-by-case}. The following questions arise naturally:  
\begin{itemize}
    \item 
    Can we provide a unified methodology for verifying  existing notions of observational properties in the literature without investigating each of them case-by-case?
    \item 
    Can we find a suitable way to classify different notions of observational properties in the literature in terms of their similarities and the verification complexity?
\end{itemize}

In this paper, we aim to answer the above two questions by providing a unified and flexible approach for verifying partially-observed DES. Our approach relies on the recently developed new temporal logic in the computer science literature called \emph{HyperLTL} \cite{clarkson2014temporal}. Particularly, HyperLTL generalized the standard linear-time temporal logic (LTL), which is evaluated over only a single trace,  by adding quantifiers among different traces. HyperLTL has been shown as a very suitable tool for expressing information-flow properties (also called hyper-properties \cite{clarkson2010hyperproperties}) in the context of formal verification. 
Specifically, our uniform framework consists of two steps. 
First, for a DES plant model, we construct the corresponding (modified) \emph{Kripke structure} that tracks both the state information and the observation information in the system. The issue of unobservability is  effectively handled by the proposed structure.  
Next, we show that most of the observational properties in the DES literature can be captured by explicit HyperLTL formulae over the constructed Kripke structure.
These properties include, but not restricted to, diagnosability, predictability, (strong/weak/I-/delayed) detectability and (initial-state/current-state/infinite-step) opacity. 

Although verification algorithms already exist for these observational properties in the literature,  our unified approach is still of significance in threefold:  
\begin{itemize}
    \item 
    First, our approach is uniform in the sense that all properties are expressed using the same logic over the same Kripke structure. 
    As a consequence, one does not need to develop a  customized verification algorithm for each property case-by-case any more. 
    \item 
    Second, by expressing observational properties of DES in terms of HyperLTL, 
    the proposed unified framework  provides the access to HyperLTL model checking algorithms for the property verification, based on which one can leverage many highly optimized efficient tools   such as \texttt{MCHyper} \cite{finkbeiner2015algorithms} and \texttt{HyperQube} \cite{hsu2021bounded}, where symbolic and bounded techniques are used. 
    \item 
    Finally, by writing down each observational property explicitly in HyperLTL, our framework  naturally provides a complexity hierarchy for different properties in terms of  the alternation depth of the quantifiers.
\end{itemize}

We would like to remark that, although HyperLTL itself is a tool for specifying information-flow properties, it cannot be directly applied to check observational properties in DES due to the following two discrepancies. 
The first technical challenge is the presence of \emph{unobservable events}. Specifically, observational properties are evaluated over the observation sequence to  which an unobservable event does not contribute. This is different from the standard HyperLTL model checking where the time-indices of the internal trace and its information-flow are the same.  Second, the semantics of HyperLTL are defined over \emph{infinite} traces while  observational properties in DES are usually concerned with \emph{finite} strings.  For example, although initial-state opacity has been expressed using HyperLTL (not for DES models and without unobservable events) \cite{wang2020hyperproperties, anand2021formal,liu2022secure}, it has been pointed out by \cite{liu2022secure} that expressing  current-state opacity or  infinite-step opacity in terms of HyperLTL is technically challenging due to the fact that the quantification acts at the beginning of trajectories rather than every instant of trajectories.  
All these technical challenges in applying HyperLTL to DES have been addressed in our results.

Finally, we note that model checking techniques have already been used in the literature for the verification of partially-observed DES. For example,  model checking for diagnosability of DES is studied in \cite{boussif2015diagnosability,tuxi2022diagnosability,pencole2022diagnosability}. However, these works still use model checking over \emph{single trace} such as LTL model checking. In order to capture the \emph{system-wide} requirements in observational properties, existing works need to build the information structure for the underlying specific property such as twin-plant for diagnosability. However, here we use HyperLTL directly, which does not need to construct an information-synchronization structure for each specific property. This information is handled directly and implicitly by the observation equivalence condition in HyperLTL formulae.

\subsection{Organization}	
The remaining part of the paper is organized as follows. In Section \ref{sec-pre}, we present the system model to be analyzed and review some necessary preliminaries on HyperLTL. In Section \ref{sec-KS}, the Kripke structure for partially-observed DES is defined. In Sections \ref{sec-diagpred}--\ref{sec-opacity}, we show explicitly how 
the notions of diagnosability, predictability, different variants of detectability and opacity can be expressed in terms of HyperLTL formulae.  Finally, we conclude the paper in Section \ref{sec-conclusion}.

\section{Preliminaries}\label{sec-pre}
\subsection{System Model}
Let $\Sigma$ be a finite set of events (or alphabets).  
A finite (respectively, infinite) string $s=\sigma_1\cdots\sigma_n(\cdots),~\sigma_i\in\Sigma$ is a finite (respectively, infinite) sequence of events. 
We denote by $\Sigma^*$ and $\Sigma^\omega$ the sets of all finite and infinite strings over $\Sigma$, respectively. The empty string $\epsilon$ is included in $\Sigma^*$. 
The length of a finite string $s\in\Sigma^*$ is denoted by $|s|$ and  with $|\epsilon|=0$. 
A $*$-language (respectively, $\omega$-language)  is a set of finite (respectively, infinite) strings. 
Given a   language $L\subseteq \Sigma^*\cup\Sigma^\omega$, the \emph{prefix closure} of language $L$, denoted by $\overline{L}$,  is defined as the set of all its finite prefixes, i.e., $\overline{L}=\{s\in\Sigma^*:\exists w\in\Sigma^*\cup\Sigma^\omega \text{ s.t. }s w\in L\}$. 
A $*$-language $L\subseteq \Sigma^*$ is said to be prefix closed if $L=\overline{L}$. 
Given any string $s\in L$, the post-language of $s$ in $L$ is defined as $L/s=\{w\in\Sigma^*\cup \Sigma^\omega:s w\in L\}$. 
	
We consider a DES modeled by a  finite-state automaton (FSA) 
\[
	G=(X,\Sigma,\delta,X_0),  
\]
where 
    $X$ is a finite set of states; 
	$\Sigma$ is a finite set of events; 
	$\delta:X\times\Sigma\to X$ is a partial transition function such that: for any $x,x'\in X$ and   $\sigma\in\Sigma$, $x'=\delta(x,\sigma)$ means that there exists a transition from state $x$ to state $x'$ with event label $\sigma$; 
and  	$X_0\subseteq X$ is the set of all possible initial states. 
We also extend the transition function to $\delta:X\times \Sigma^*\to X$ recursively by: 
	(i)  $\delta(x,\epsilon)=x$; and 
	(ii) for any $x\in X$, $s\in \Sigma^*$, and $\sigma\in\Sigma$, we have $\delta(x,s\sigma)=\delta(\delta(x,s),\sigma)$. 
We define $\L(G,x)=\{s\in\Sigma^*:\delta(x,s)!\}$ as the set of finite strings that can be generated by system $G$ from state $x\in X$. For simplicity, we define $\L(G)=\cup_{x_0\in X_0}\L(G,x_0)$ as the $*$-language generated by system $G$. 
Similarly, we define $\L^\omega(G,x)=\{s\in\Sigma^\omega: \forall{t}\in \overline{\{s\}}\text{ s.t. }\delta(x,t)!\}$ as the infinite strings generated by system $G$ starting from $x\in X$ and we also define $\L^\omega(G)=\cup_{x_0\in X_0} \L^\omega(G,x_0)$. 
	
In a partially-observed DES, not all events can be observed perfectly. 
To this end, we consider an observation mask function 
\[
	M:\Sigma\to  \O \cup\{\epsilon\},
\]
where $\O$ is the set of observation symbols. 
An event $\sigma\in\Sigma$ is said to be \emph{unobservable} if $M(\sigma)=\epsilon$; otherwise, it is observable. 
We denote by $\Sigma_o$ and $\Sigma_{uo}$ the sets of observable and unobservable events, respectively. 
Moreover, events $\sigma,\sigma'\in\Sigma$ are said to be \emph{indistinguishable} if $M(\sigma)=M(\sigma')$. The mask function is also extended to $M:\Sigma^*\cup\Sigma^\omega\to \O^*\cup\O^\omega$ such that, for any $s\in \L(G)$, we have
	\begin{enumerate}[(i)]
	    \item 
	    $M(\epsilon)=\epsilon$; and 
	    \item 
	    for any $s\in\Sigma^*,\sigma\in\Sigma$, we have 
	    $M(s\sigma)=M(s)M(\sigma)$.
	\end{enumerate}
We also extend mask to $M\!:\!2^{\Sigma^*\cup\Sigma^\omega}\!\to\! 2^{\O^*\cup\O^\omega}$ by: 
	for any $L\subseteq \Sigma^*\cup\Sigma^\omega\!:\! M(L)\!=\!\{M(s)\!:\! s\!\in\! L\}$. 
	Therefore, $M(\L(G))$ and $M(\L^\omega(G))$ are the observed $*$- and $\omega$-languages generated by system $G$, respectively.
	
    For simplicity, we make the following standard assumptions in the analysis of  partially-observed DES:
    \begin{enumerate}
        \item [A1] System $G$ is live, i.e., $\forall x\in X,\exists\sigma\in\Sigma:\delta(x,\sigma)!$; and 
        \item [A2] System $G$ does not contain an unobservable cycle, i.e., $\forall x\in X, \forall s\in\Sigma^* \setminus \{\epsilon\}:x=\delta(x,s)\Rightarrow M(s)\neq \epsilon$.
    \end{enumerate}
    
Since the system is partially-observed, the system-user needs to determine the state of the system based on the observation string, which is referred to as the \emph{state estimation problem}. In this paper, we will consider the following three types of state estimates. All   properties of partially-observed DES in this paper will be defined using state estimates  \cite{yin2019estimation}.

\begin{mydef}[\bf State Estimates]\upshape
Let   $\alpha\! \in\! M(\L(G))$ be an observation string. Then
	\begin{itemize}
	\item the \emph{initial-state estimate} upon  observation $\alpha$ is the set of initial states the system could start from initially, i.e.,
	    \begin{equation}\label{eq-ise}
	     \!\!\hat{X}_{G,0}(\alpha)\!=\!\{x_0\!\in\! X_0:  
	        \exists s\!\in\! \L(G,x_0)\text{ s.t. } M(s)\!=\!\alpha\}
	    \end{equation}
	    \item the \emph{current-state estimate} upon  observation $\alpha$  is the set of states the system could be in currently, i.e.,
	   \begin{align}\label{eq-cse} 
	          \!\! \!\!   \hat{X}_G(\alpha)\!=\!
	            \left\{\delta(x_0,s) \!\in\! X:\!\!\!\!\!\!\!\!\!
	            \begin{array}{cc}
	                 &    \exists x_0\!\in\! X_0,   s\!\in\! \L(G,x_0) \\
	       &\text{ s.t. }    M(s)\!=\!\alpha 
	            \end{array} \!\!\!
	   \right\} 
	    \end{align}
	    \end{itemize}
Furthermore, let $\alpha\beta \in M(\L(G))$ be an observation string, where $\alpha$ is a prefix of the entire observation $\alpha\beta$. Then
	    \begin{itemize}
	    \item the \emph{delayed-state estimate} 
	    for the instant of $\alpha$ upon observation $\alpha\beta$ is   the set of states the system could be in $|\beta|$ steps ago when $\alpha\beta$ is observed, i.e.,
	    \begin{align}\label{eq-dse} 
	           \!\! \!\!\!\hat{X}_G(\alpha&\mid\alpha\beta)\!=\!
	   \left\{\! \delta(x_0,s)\!\in\! X \! \! :\!\!\!\!\!\!\!\!\!\!
	   \begin{array}{cc}
	    & \exists x_0\!\in\! X_0,sw\!\in\!\L(G,x_0) \text{ s.t.}\\
	    &M(s)\!=\!\alpha\wedge M(sw)\!=\!\alpha\beta
	   \end{array} \!\!\!
	    \right\} 
	    \end{align}
	\end{itemize}    
\end{mydef}

\subsection{LTL and HyperLTL} 
Let $\mathcal{AP}$ be a set of atomic propositions representing some basic properties of interest. A \emph{trace}  $\pi=\pi_0\pi_1\cdots\in (2^\mathcal{AP})^\omega$ is an infinite sequence over $2^\mathcal{AP}$.  
We denote by $\pi[i]=\pi_i$ the $i$th element in the trace and by  $\pi[i,\infty]= \pi_i\pi_{i+1}\cdots\in (2^{\mathcal{AP}})^\omega$ its suffix starting from the $i$th instant. 
Linear Temporal Logic (LTL) is a widely used approach for  evaluating whether or not a trace $\pi$ satisfies some desired property.  
The syntax of LTL formulae is as follows
\[
\psi ::=a \mid\neg\psi\mid\psi\vee\psi\mid\bigcirc\psi\mid\psi \U \psi,  
\]
where  $a\in \mathcal{AP}$ is an  atomic proposition;  
$\neg$ and $\vee$ are Boolean operators ``negation" and ``disjunction", respectively, and  $\bigcirc$ and $\U$ are temporal operators  ``next" and ``until", respectively. 
We can also define more Boolean operators such as  ``implication" by $\psi_1\to\psi_2\equiv\neg\psi_1\wedge\psi_2$ and ``conjunction" by $\psi_1\wedge\psi_2\equiv\neg(\neg\psi_1\vee \neg\psi_2)$.  
Furthermore, we can induce temporal operators  ``eventually" by  $\lozenge\psi\equiv \top \U \psi$  and ``always" by $\square\psi\equiv \neg \lozenge\neg \psi$.  
The semantics of LTL can be found in \cite{baier2008principles}. 
We denote by $\pi\models \varphi$ when trace $\pi$ satisfies LTL formula $\varphi$.  

Note that LTL can only evaluate the correctness of a \emph{single trace}. In many applications, however, the desired property is \emph{system-wide} and can only be evaluated among multiple traces. For example, in diagnosability analysis, we need to check the existence of two strings with the same observation: one is fault but the other is normal. Then HyperLTL generalizes LTL by further supporting \emph{trace quantifiers}. 
Formally, let $\mathcal{V}=\{\pi_1,\pi_2,\ldots\}$ be the set of \emph{trace variables}, where each   $\pi_i$ represents an individual trace. Then,  the syntax of HyperLTL  formulae is as follows \cite{clarkson2014temporal}:
	\begin{flalign}
		\phi&::=\exists\pi.~\phi\mid\forall\pi.~\phi\mid\psi,\notag\\
		\psi&::=a^\pi\mid\neg\psi\mid\psi\vee\psi\mid\bigcirc\psi\mid\psi \U \psi,\notag
	\end{flalign}
where $\exists$ and $\forall$ are  the universal and the existential trace quantifiers, representing ``for some trace" and ``for all traces", respectively. 
Formula $\psi$ is just an LTL formula except that the atomic propositions can refer to distinct trace variables. Particularly, since HyperLTL formulae can refer to multiple traces, we denote by $a^\pi$ an atomic proposition $a\in \mathcal{AP}$ that should be checked on trace $\pi$. 
	
The semantics of  HyperLTL are defined over \emph{a set of}  traces $T\!\subseteq\!(2^{\mathcal{AP}})^\omega$ 
and  
a \emph{partial} mapping (called trace assignment) $\Pi\!:\!\mathcal{V}\!\to\! (2^{\mathcal{AP}})^\omega$.  
In particular, we denote by $\Pi_\emptyset$ the empty assignment whose domain is the empty set $\emptyset$.
We denote by $\Pi[ \pi\mapsto \xi]$ the same trace assignment as $\Pi$ expect that $\pi$ is mapped to $\xi$. 
The trace assignment \emph{suffix} $\Pi[i,\infty]$ denotes the trace assignment $\Pi'(\pi)=\Pi(\pi)[i,\infty]$ for all $\pi$.
Then we denote by $(T,\Pi) \models \phi$ that HyperLTL $\phi$ is satisfied over a set of traces $T\subseteq(2^{\mathcal{AP}})^\omega$ and trace assignment  $\Pi:\mathcal{V}\to (2^{\mathcal{AP}})^\omega$, which is defined as follows: 
\begin{equation}
\begin{tabular}{lcl}
$(T,\Pi)\models \exists\pi.\phi$  &  iff  & $\exists \xi\in T:(T,\Pi[\pi\mapsto \xi])\models\phi$
\\
$(T,\Pi)\models\forall\pi.\phi$  &  iff  & $\forall \xi\in T:(T,\Pi[\pi\mapsto \xi])\models\phi$
\\
$(T,\Pi)\models a^\pi$ &  iff   &  $a\in\Pi(\pi)[0]$
\\
$(T,\Pi)\models\neg\psi$ &  iff  & $(T,\Pi) \nvDash\psi$
\\
$(T,\Pi)\models\psi_1\vee\psi_2$ &  iff  & $(T,\Pi)\models\psi_1\text{ or }(T,\Pi)\models\psi_2$
\\
$(T,\Pi)\models \bigcirc\psi$ &  iff  & $(T,\Pi[1,\infty])\models\psi$
\\
$(T,\Pi)\models \psi_1 \U\psi_2$ &  iff  & $\left(\exists i\geq 0:(T,\Pi[i,\infty])\models\psi_2\right)\wedge$
\\
&&$\left(\forall 0\leq j<i:(T,\Pi[j,\infty])\models\psi_1\right)$\notag
\end{tabular}
\end{equation}
We say   a set of traces $T\subseteq(2^{\mathcal{AP}})^\omega$ satisfy a HyperLTL formual $\phi$, denoted by 
$T \models \phi $, if   $(T,\Pi_\emptyset) \models \phi$. 
    
\subsection{Kripke Structure}
In model checking of HyperLTL, the set of traces $T\subseteq(2^{\mathcal{AP}})^\omega$ are usually generated by a  \emph{Kripke structure}. Formally, a Kripke structure is a tuple $K=(Q,Q_0,\Delta,\mathcal{AP},L)$ where $Q$ is the set of states, $Q_0\subseteq Q$ is the set of initial states, $\Delta \subseteq Q\times Q$ is the   transition relation, $\mathcal{AP}$ is the set of atomic propositions  and $L:Q\to 2^\mathcal{AP}$ is the labeling function. 
	
We say $\rho=\rho_0\rho_1\ldots\in Q^\omega$  is a \emph{run} in $K$ if  $\rho_0\in Q_0$ and   $\langle \rho_i, \rho_{i+1}\rangle \in \Delta , \forall i\geq 0$.   
We say $\pi=\pi_0\pi_1\ldots\in (2^{\mathcal{AP}})^\omega$  is a \emph{trace} in $K$ if  there exists a run $\rho=\rho_0\rho_1\ldots\in Q^\omega$ such that $\pi_i=L(\rho_i),\forall i\geq 0$. 
We denote by $\run(K)$ and $\trace(K)$ the set of all runs and traces generated by $K$, respectively.  Then we say that a Kripke structure $K$ satisfies HyperLTL formula $\phi$, denoted by $K\models\phi$, if $\trace(K)\models \phi$.
	
\section{Partially-Observed DES in Kripke Structure}\label{sec-KS}
The main objective of this paper is to use HyperLTL model checking techniques to solve the observational property verification problems for  partially-observed DES. 
To this end, we need to transform the FSA model for DES into a Kripke structure for the purpose of model checking.

In HyperLTL model checking, atomic propositions are usually assigned to each state or transition in the system model. However, in the setting of partially-observed DES, we note that, for any internal string $s=\sigma_1\sigma_2\cdots\in \Sigma^\omega$, its observation string 
is $M(\alpha)=o_1o_2\cdots\in \mathcal{O}^\omega$, where for each $i\geq 1$, $o_i$ is not necessarily the observation of event $\sigma_i$, since there may have unobservable strings in between.  
Therefore, the time-indices of the internal string and its information-flow may be mismatched. 
We also cannot assign the empty proposition to those unobservable transitions  since it means ``no property of interest",  which is different from the empty observation. 
	
To address the issue of unobservability, let $x,x'\in X$ be two states in $G$ and 
$o\in \Delta\cup \{\epsilon\}$ be an observation symbol including the empty observation $\epsilon$. 
We denote by $x \stackrqarrow{o}{}x' $ if $x$ can reach $x'$ via some string whose observation is $o$, i.e., $\exists s\in \mathcal{L}(G,x)$ such that $\delta(x,s)=x'$ and $M(s)=o$.  
Note that, without unobservable event, the above string $s$ must be a single event when $o\in \Delta$, 
and  be $\epsilon$ when $o=\epsilon$. However, for the general case, $s$ can be a string with more than one event even when $o=\epsilon$.

	\begin{figure}
	    \centering
		\subfigure[System $G$.]{
			\centering
			\begin{tikzpicture}[->,>={Latex}, thick, initial text={}, node distance=1.8cm, initial where=left, thick, base node/.style={circle, draw, minimum size=7mm, font=\normalsize}]  
			\node[state, initial, base node, ] (0) {$x_0$};
			\node[ ] at ($(0.center)+(1.3cm,0cm)$) (1)  {$\cdots$};
			\node[state, base node, ] [right of=1]  (2) {$x$};
			\node[state, base node, ] [right of=2] (3) {$x'$};
			\node[state, base node, ] [below of=0] (4) {$x_1$};
			\node[ ] at ($(4.center)+(1.3cm,0cm)$) (5)  {$\cdots$};
			\node[state, base node, ] [right of=5] (6) {$y$};
			\node[state, base node, ] [right of=6] (7) {$y'$};
			\node[ ] at ($(5.center)+(0.9cm,0cm)$) (8)  {$\cdots$};
			\node[ ] at ($(3.center)+(0.8cm,0cm)$) (9)  {$\cdots$};
			\node[ ] at ($(7.center)+(0.8cm,0cm)$) (10)  {$\cdots$};
			
			\path[->]
			(0) edge node  {} (1)
			(1) edge node [yshift=0.2cm] {\normalsize $\sigma/o$} (2)
			(2) edge node [yshift=0.2cm] {\normalsize $\sigma'/o'$} (3)
			(0) edge node [xshift=0.33cm] {\normalsize $u/\epsilon$} (4)
			(4) edge node  {} (5)
			(3) edge node [xshift=-0.2cm, yshift=0.2cm] {\normalsize $u/\epsilon$} (6)
			(6) edge node [yshift=0.2cm] {\normalsize $\sigma''/o''$} (7)
			;
			
			\end{tikzpicture}
		}
		\\
		\subfigure[Kripke structure $K_G$.]{
			\centering
			\begin{tikzpicture}[->,>={Latex}, thick, initial text={}, node distance=1.9cm, initial where=left, thick, base node/.style={rectangle, rounded corners, draw, minimum size=7mm, font=\normalsize}]  
			\node[initial, state, base node,  ] (0) {$(x_0,\epsilon)$} ;
			\node[ ] at ($(0.center)+(1.6cm,0cm)$) (1)  {$\cdots$};
			\node[state, base node, ] at ($(1.center)+(1.6cm,0cm)$) (2) {$\,(x,o)\,$};
			\node[state, base node, ] [right of=2] (3) {$(x',o')$};
			\node[initial, state, base node, ] [below=1.6cm] (4) {$(x_1,\epsilon)$};
			\node[ ] at ($(4.center)+(1.6cm,0cm)$) (5)  {$\cdots$};
			\node[state, base node, ] at ($(5.center)+(1.6cm,0cm)$) (6) {$\,(y,o')\,$};
			\node[state, base node, ] [right of=6] (7) {$(y',o'')$};
			\node[ ] at ($(5.center)+(0.5cm,0cm)$) (8)  {$\cdots$};
			\node[ ] at ($(3.center)+(1.1cm,0cm)$) (9)  {$\cdots$};
			\node[ ] at ($(7.center)+(1.1cm,0cm)$) (10)  {$\cdots$};
			
			\path[->]
			(0) edge node  {} (1)
			(1) edge node [yshift=0.2cm] {} (2)
			(2) edge node [yshift=0.2cm] {} (3)
			(0) edge node [xshift=0.33cm] {} (4)
			(4) edge node  {} (5)
			(6) edge node [yshift=0.2cm] {} (7)
			(2) edge node [yshift=0.2cm] {} (6)
			(3) edge node [yshift=0.2cm] {} (7)
			;
			
			\node[font=\normalsize ] at ($(0.center)+(0.5cm,0.6cm)$) (0label)  {$\{x_0\}$};
			\node[font=\normalsize ] at ($(4.center)+(0.5cm,0.6cm)$) (4label)  {$\{x_1\}$};
			\node[font=\normalsize ] at ($(2.center)+(0.65cm,0.6cm)$) (2label)  {$\{x,o\}$};
			\node[font=\normalsize ] at ($(3.center)+(0.65cm,0.6cm)$) (3label)  {$\{x',o'\}$};
			\node[font=\normalsize ] at ($(6.center)+(0.7cm,0.6cm)$) (6label)  {$\{y,o'\}$};
			\node[font=\normalsize ] at ($(7.center)+(0.7cm,0.6cm)$) (7label)  {$\{y',o''\}$};
			
			\end{tikzpicture}
		}
		\caption{Conceptual illustration of how to construct $K_G$ from $G$.}\label{fig-transition1}
	\end{figure}
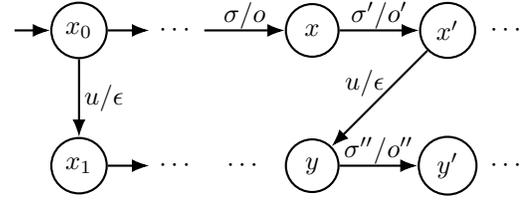
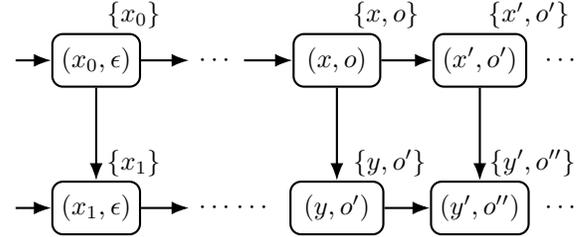

Now, we present how to construct the Kripke structure associated with a DES for the purpose of verification of observational properties. 
\begin{mydef}[\bf Kripke Structure for DES]\upshape
Given partially-observed DES $G$ with mask $M:\Sigma\to  \O \cup\{\epsilon\}$, 
its associated Kripke structure is defined by: 
    \[
    K_G=(Q,Q_0,\Delta,\AP,L)
    \]
    where
    \begin{itemize}
    \item 
    $Q\subseteq X \times  (\O \cup \{\epsilon\})$ is the set of states;
    \item 
    $Q_0= \{(x,\epsilon) \in X\times \{\epsilon\}  :  \exists x_0\in X_0\text{ s.t. } x_0 \stackrqarrow{\epsilon}{}x\}$  is the set of initial states;
    \item 
    $\Delta\subseteq Q\times Q$ is the transition function  defined by: 
    for any two  states $q=(x,o),q'=(x',o')\in   Q $, 
    we have 
    \[
    \langle (x, o) , (x', o')  \rangle \in \Delta  
    \text{ iff  } x \stackrqarrow{o'}{}x'  \wedge o'\in \mathcal{O}
    \]
  \item 
  $\AP=X\cup\O$ is the set of atomic propositions;
  \item 
  $L:Q\to 2^{\AP}$ is the labeling function   defined by: 
  for each $q=(x,o)\in Q$, we have 
  \begin{align} \label{eq-label}
		L(q)= 
		\left\{  
		\begin{array}{l l}
			\{  x \}    &\text{if }    q\in Q_0 \\ 
			\{ x,o \}   &\text{if }    q\notin Q_0 
		\end{array} 
		\right..
	\end{align} 
\end{itemize}
\end{mydef}

A conceptual illustration of how to construct $K_G$ from $G$ is shown in Figure \ref{fig-transition1}.
Intuitively, for each state $(x,o)$ in   $K_G$, 
the first component $x$  captures  the current state of $G$  and the second component $o$ captures 
the  latest observation symbol happened in $G$.   
At each state $(x,o)$, when a feasible string $s\in \L(G,x)$ 
such that $M(s)\!=\!o'$ occurs, the Kripke structure moves to new state $(\delta(x,s) , o')$, where the first component is determined by the transition function in $G$ and the second component simply records the observed symbol.  
Note that the above string $s$ may be in the form of $s\!=\!w_1\sigma w_2$, where $w_1,w_2\!\in\! \Sigma_{uo}^*$ and $M(\sigma)\!=\!o$. 
Since there is no event observed initially, the initial state is of form $(x,\epsilon)$. 
The labeling function assigns the state symbol $x$ and the observation symbol $o$ as the atomic propositions hold at each $(x,o)$.  
Therefore, the trace information in $K_G$   already contains both the state sequence information and the observation sequence information, which are sufficient for the purpose of  verifying  observational properties of $G$.

To formally see the connection between $G$ and $K_G$, 
let us consider an arbitrary infinite string $s\in \L^\omega(G,x_0)$ in $G$. Note that, we can always write $s$ in the form of 
\[
s= w_0\sigma_1w_1\sigma_2w_2 \cdots \in \Sigma^\omega
\]
where each  $w_i\in \Sigma_{uo}^*$ is an unobservable string 
and each $\sigma_i\in \Sigma_o$ is an observable event with $M(\sigma_i)=o_i$. 
Let 
\begin{equation}\label{eq:fragment}
   \underbrace{x_0^0 \cdots x_0^{|w_0|}}_{\text{visited along }w_0}
\underbrace{x_1^0 \cdots x_1^{|w_1|}}_{\text{visited along }\sigma_1w_1}
\underbrace{x_2^0 \cdots x_2^{|w_2|}}_{\text{visited along }\sigma_2w_2}\cdots \in X^\omega  
\end{equation}
be the infinite sequence of states visited along $s$ from $x_0$, where 
$x_0^0\!=\!x_0$. 
Note that, in the construction of $K_G$, upon each observation, we will ``jump" directly to a state without considering the states visited by unobservable strings in between. 
Therefore, we know that, 
\emph{for any} of the indices $k_0,k_1,\dots$, where  
$k_i\in \{0,\dots, |w_i|\}$, the following run exists in $K_G$
\[
\rho=(x_0^{k_0},\epsilon)(x_1^{k_1},o_1)(x_2^{k_2},o_2)\cdots\in\run(K_G)
\]
We call such a run \emph{compatible} with string $s$ from initial state $x_0$. 
Since the choices of the indices $k_0,k_1,\dots$  are not unique, 
we denote by $\run(s,x_0)\subseteq \run(K_G)$ the set of all runs that are compatible with $s$ and $x_0$. 
Note that when $s\in\L(G,x_0)$ is a finite string, there also exists a finite run that is compatible with $s$ and $x_0$. With a slight abuse of notation, we still denote this by $\rho\in\run(s,x_0)$.

On the other hand, for any run 
\[
\rho=(x_0,\epsilon)(x_1,o_1)(x_2,o_2)\cdots\in\run(K_G)
\]
by construction, we have 
$x_i \stackrqarrow{o_{i+1}}{}x_{i+1}$. 
Therefore, we can always find an initial state $\hat{x}_0\in X_0$ and an infinite string $s\in \L^\omega(G,\hat{x}_0)$ such that 
$M(s)=o_1o_2\cdots$ and  each $x_i$ is reached by a prefix of $s$ whose observation is $o_1o_2\cdots o_i$. That is,  $\rho\in \run(s,\hat{x}_0)$.

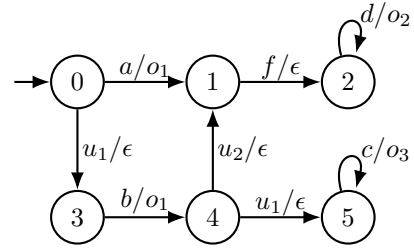
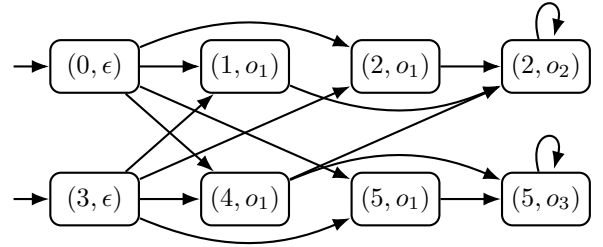
\begin{figure}
		\centering
		\subfigure[System $G$.\label{fig-exam-diagpred-G}]{
			\centering
			\begin{tikzpicture}[->,>={Latex}, thick, initial text={}, node distance=1.8cm, initial where=left, thick, base node/.style={circle, draw, minimum size=7mm, font=\normalsize}]   
			\node[state, initial, base node, ] (0) {$0$};
			\node[state, base node, ] (1) [right of=0] {$1$};
			\node[state, base node, ] (2) [right of=1] {$2$};
			\node[state, base node, ] (3) [below of=0] {$3$};
			\node[state, base node, ] (4) [right of=3] {$4$};
			\node[state, base node, ] (5) [right of=4] {$5$};

			\path[->]
			(0) edge node [yshift=0.2cm] {\normalsize $a/o_1$} (1)
			(0) edge node [xshift=0.4cm] {\normalsize $u_1/\epsilon$} (3)
			(1) edge node [yshift=0.2cm] {\normalsize $f/\epsilon$} (2)
			(2) edge [loop above ] node         [xshift=0.5cm, yshift=-0.2cm]     {\normalsize $d/o_2$} ()
			(5) edge [loop above ] node         [xshift=0.5cm, yshift=-0.2cm]     {\normalsize $c/o_3$} ()
			(3) edge node [yshift=0.23cm] {\normalsize $b/o_1$} (4)
			(4) edge node [yshift=0.2cm] {\normalsize $u_1/\epsilon$} (5)
			(4) edge node [xshift=0.4cm] {\normalsize $u_2/\epsilon$} (1);
			
			\end{tikzpicture}
		}
		\\
		\subfigure[Kripke structure $K_G$.\label{fig-exam-diagpred-KG}]{
			\centering
			\begin{tikzpicture}[->,>={Latex}, thick, initial text={}, node distance=2cm, initial where=left, thick, base node/.style={rectangle, rounded corners, draw, minimum size=7mm, font=\normalsize}]   
			\node[initial, state, base node, rectangle, rounded corners, ] (0) {$\;(0,\epsilon)\;$} ;
			\node[state, base node, rectangle, rounded corners, ] [right of=0] (1) {$(1,o_1)$} ;
			\node[state, base node, rectangle, rounded corners, ] [right of=1] (2) {$(2,o_1)$} ;
			\node[state, base node, rectangle, rounded corners, ] [right of=2] (22) {$(2,o_2)$};
			\node[initial, state, base node, rectangle, rounded corners, ] [below=1.4cm] (3)  {$\;(3,\epsilon)\;$};
			\node[state, base node, rectangle, rounded corners, ] [right of=3] (4) {$(4,o_1)$};
			\node[state, base node, rectangle, rounded corners, ] [right of=4] (5) {$(5,o_1)$};
			\node[state, base node, rectangle, rounded corners, ] [right of=5] (55) {$(5,o_3)$};
			
			
			\path[->]
			(0) edge node  {} (1)
			(0) edge node  {} (4)
			(4) edge node  {} (22)
			(2) edge node  {} (22)
			(22)   edge [loop above] node {} ()
			(3) edge node  {} (4)
			(3) edge node  {} (1)
			(3) edge node  {} (2)
			(0) edge node  {} (5)
			(5) edge node  {} (55)
			(55)   edge [loop above] node {} ()
			;
			
			\draw[->]
            (0) to [bend right=-23] (2);
            \draw[->]
            (1) to [bend right=23] (22);
			\draw[->]
            (3) to [bend right=23] (5);
            \draw[->]
            (4) to [bend right=-23] (55);
            
			
			\end{tikzpicture}
		}
		\caption{Example for a partially-observed DES and its Kripke structure.}\label{fig-exam-diag}
	\end{figure}

We illustrate the construction of Kripke structure $K_G$ from DES $G$ by the following example. 
\begin{myexm}\upshape
Let us consider system $G$ shown in Figure \ref{fig-exam-diagpred-G}, 
where $X\!=\!\{0,1,2,3,4,5\}$, $\Sigma_o\!=\!\{a,b,c,d\}$, $\Sigma_{uo}\!=\!\{u_1,u_2,f\}$,  $\O=\{o_1,o_2,o_3\}$  and the observation mask 
  $M:\Sigma \to \O$ is defined by: $M(a)\!=\!M(b)\!=\!o_1,M(c)\!=\!o_3,M(d)\!=\!o_2$ and $M(u_1)\!=\!M(u_2)\!=\!M(f)\!=\!\epsilon$. 
The initial states of $K_G$ are $(0,\epsilon)$ and $(3,\epsilon)$ since the system may reach state $3$ from the initial state $0$ via unobservable string $u_1$. From state $(0,\epsilon)$, by observing symbol $o_1$, one may reach states $1,2,4,5$. 
Therefore, transitions from $(0,\epsilon)$ to states $(1,o_1),(2,o_1),(4,o_1)$ and $(5,o_1)$ are all defined in $K_G$.  The labeling function can be encoded directly from the state,  e.g., $L((0,\epsilon))\!=\!\{0\}$
and $L((1,o_1))\!=\!\{1,o_1\}$.   
For example, let us consider initial-state $x_0=0$ and infinite string 
$s\!=\!u_1 b u_2 f (d)^\omega \!\in\! \mathcal{L}^\omega(G)$ with $M(s)\!=\!o_1(o_2)^\omega$. 
Then  a run in $K_G$ compatible with $s$  and  $x_0$ can be, e.g., 
$(0,\epsilon)(4,o_1)(2,o_2)^\omega \!\in\! \run(s,x_0)$ 
or
$(3,\epsilon)(1,o_1)(2,o_2)^\omega \!\in\! \run(s,x_0)$. 
\end{myexm}

\section{Diagnosability \& Predictability in HyperLTL}\label{sec-diagpred}
In this section, we consider the verification of diagnosability and predictability in partially-observed DES. 
In this setting, it is assumed that system $G$ may contain some faults modeled as a set of fault events $\Sigma_\F \subseteq \Sigma$. 
With a slight abuse of notation, for any string $s\in \Sigma\cup \Sigma^\omega$, we denote by $\Sigma_\F\in s$ if string $s$ contains a fault event in $\Sigma_\F$. 
For simplicity, we assume that the state-space of $G$ is partitioned as 
\[
X=X_\N\dot{\cup}X_\F, 
\]
where $X_\N$ is the set of normal states and $X_\F$ is the set of fault states such that 
\begin{equation}\label{eq-partition}
    \forall x_0\in X_0, \forall s\in \mathcal{L}(G,x_0): \Sigma_\F\in s\Leftrightarrow \delta(x_0,s)\in X_\F.
\end{equation}
Note that this assumption is without loss of generality, since we can always refine the state-space of $G$ such that the partition holds.  

Diagnosability then characterizes whether or not we can always detect the occurrence of a fault event within a finite number of steps. 
To this end, we define 
\[
\Psi_\F=\{s\in\L(G): \Sigma_\F\in s\wedge (\forall t\in \overline{\{s\}}\setminus \{s\}: \Sigma_\F\notin t)     \}
\] as the set of strings in which fault event occurs for the first time. The notion of diagnosability is reviewed as follows \cite{sampath1995diagnosability}. 
\begin{mydef}[\bf Diagnosability]\label{def-diag}\upshape
Given system $G$, observation mask $M:\Delta\to\O\cup\{\epsilon\}$  and fault events $\Sigma_\F\subseteq \Sigma$, 
we say system $G$ is \emph{diagnosable} if any occurrence of fault can always be determined within a finite number of delays, i.e.,
	\begin{flalign}
	  (\exists n\in\NN)(\forall &s\in\Psi_\F)(\forall t\in\L(G)/s)\notag\\
	  &[|M(t)|\geq n\Rightarrow\hat{X}_G(M(st))\subseteq X_\F].
	\end{flalign}
\end{mydef}

When it refers to fault prediction,  the objective is to predict the occurrences of fault events \emph{in advance} such that 
(i)  there is no miss-alarm in the sense that any fault can be alarmed before it actually occurs; and 
(ii) there is no false-alarm in the sense that, once a fault alarm is issued, fault events will occur inevitably within a finite number of step. 
To define the notion of predictability, it is convenient to define
	\begin{itemize}
	    \item 
	    the set of \emph{boundary states} $\partial (G)$, which is the set of normal states from which  a fault event can occur in the next step, i.e., 
	    \[\partial (G)=\left\{x\in X_\N:\exists\sigma\in\Sigma_\F\text{ s.t. }\delta(x,\sigma)!\right\}\]
	    \item 
	    the set of \emph{indicator states} $ \I(G)$, which is the set of normal states from which the system will enter fault states \emph{inevitably} within a finite number of steps, i.e, 
	    \[
	    \I(G)=\left\{x\in X_\N:\!\!\!\!\!\!\!
	    \begin{array}{cc}
	    &\exists n\in\NN,\forall s\in\L(G,x)\\
	    &\text{s.t. }|s|\geq n\Rightarrow\delta(x,s)\in X_\F
	    \end{array}\!\!
	    \right\}
	    \]
	\end{itemize}
 Using the notions of boundary states and indicator states, we recall the definition of predictability as follows \cite{genc2009predictability}.
\begin{mydef}[\bf Predictability]\label{def-pred}\upshape
Given system $G$, observation mask $M:\Delta\to\O\cup\{\epsilon\}$ and fault events $\Sigma_\F\subseteq \Sigma$, 
we say system $G$ is \emph{predictable} if the occurrence of fault can always be alarmed before it happens, i.e.,
	\begin{flalign}
	&(\forall x_0\in X_0)(\forall s\in\L(G,x_0) :\delta(x_0,s)\in \partial(G))\notag\\
	&~~~~~~~~~~~~~~~~~~~~~~~~(\exists t\in\overline{\{s\}})[ \hat{X}_G(M(t))\subseteq \I(G)].
	\end{flalign}
	\end{mydef}

\subsection{Diagnosability in HyperLTL}

Before expressing diagnosability in HyperLTL, we introduce some notation simplifications for HyperLTL formulae. 
For any trace variable $\pi$, we define 
$\F^\pi$  as the proposition that the trace is at a fault state, i.e.,  $\F^\pi\equiv \bigvee_{x\in X_\F}x^{\pi}$. 
Also, let $\pi_1$ and $\pi_2$ be two trace variables. Then we define
\[
o^{\pi_1}=o^{\pi_2}\text{ iff }\bigwedge_{o\in \O}o^{\pi_1}\leftrightarrow o^{\pi_2}
\]
which is the proposition that $\pi_1$ and $\pi_2$ are observational-equivalent \emph{at the initial instant}. As such, formula $\square(o^{\pi_1}=o^{\pi_2})$ represents that two infinite traces $\pi_1$ and $\pi_2$ are observational-equivalent at any instants.

Now, we present the following main theorem, stating how to formulate diagnosability of $G$ using HyperLTL for Kripke structure $K_G$.

    \begin{tcolorbox} 
        \begin{mythm}[\textbf{HyperLTL for Diagnosability}]\label{thm-diag}\upshape  
        System $G$ is diagnosable if and only if $K_G\models\phi_{dia}$,
        where
        \begin{equation}
        \phi_{dia}=\forall\pi_1.\forall\pi_2.\left[\lozenge \F^{\pi_1}\wedge\square(o^{\pi_1}=o^{\pi_2})\to\lozenge\F^{\pi_2}\right].
        \end{equation}
        \end{mythm}
    \end{tcolorbox} 
    
    Intuitively, the above theorem says that, to make the system diagnosable, for any two infinite strings having the same observation, if one string contains a fault event, i.e., $\lozenge \F^{\pi_1}$, then the other string should also contain a fault event, i.e., $\lozenge \F^{\pi_2}$. Otherwise, if $\square \neg \F^{\pi_2}$, since $\square(o^{\pi_1}=o^{\pi_2})$, then the fault in the former string can never be determined within any finite number of steps.  Following this intuition, we present the formal proof of Theorem \ref{thm-diag}.
    
    \begin{proof}
    ($\Rightarrow$)
    For the sake of contraposition, suppose that $K_G\!\not\models\!\phi_{dia}$, i.e., 
    there are two runs
    \[
    \rho^i=(x_0^i,\epsilon)(x_1^i,o_1^i)\cdots\in\run(K_G),i=1,2
    \]
    such that 
    (i) $o_j^1=o_j^2=:o_j,\forall j\geq 0$; 
    (ii) $x_j^2\notin X_\F,\forall j\geq 0$; and 
    (iii) 
    for some $k\geq 1$, we have $x_k^1\in X_\F$ and $x_j^1\notin X_\F,\forall j<k$.   
    Based on the construction of $K_G$, 
    for each $i=1,2$, there exists an initial-state $\hat{x}_0^i$ and string $s^i\in \mathcal{L}^\omega(G,\hat{x}_0^i)$ such that 
    $\rho^i\in \run(s^i,\hat{x}_0^i)$. 
    Let $s_F\in \overline{\{s^1\} }$ be the prefix of $s^1$ such that $\delta(\hat{x}_0^1,s_F)=x_k^1$.
    Without loss of generality, we assume that $s_F\in \Psi_\F$; otherwise, we can use $\delta(\hat{x}_0^1,s_F)$ to replace $x_k^1$ in run $\rho^1$.   
    However, since $M(s^1)=M(s^2)=o_1o_2\cdots$ and $\Sigma_\F\notin s_2$, 
    for any $n\in \NN$, 
    we can choose $s_F\in \Psi_\F$ and $s_F t\in  \overline{\{s^1\} }$ such that 
    $|M(t)|>n$. 
    Then we have $\{\delta(x_0^1,s_Ft),\delta(x_0^2,s_N)\}\subseteq \hat{X}_G(M(st))\not\subseteq X_\F$, 
    where  $s_N\in  \overline{\{s^2\} }$ is a prefix of $s^2$ such that $\Sigma_\F\notin s_N$ and $M(s_N)=M(s_Ft)$.     This shows that $G$ is not diagnosable.

($\Leftarrow$)
We still prove this direction by contraposition. 
Suppose that $G$ is not diagnosable, i.e., 
for any $n\in\NN$, we can find strings $s\in\Psi_\F$ and $t\in\L(G)/s$
such that $|M(t)|\geq n$ but $\hat{X}_G(M(st))\not\subseteq X_\F$. 
Let us consider the case of $n\!>\!|X|^2\!+\!1$. 
Since $s\!\in\!\Psi_\F$, we have $\delta(x_0^1,s)\!\in\! X_\F$ for some $x_0^1\!\in\! X_0$. 
We denote $M(s)\!=\!o_1o_2\cdots o_k$ and $x_k^1\!=\!\delta(x_0^1,s)$.
Since $\hat{X}_G(M(st))\not\subseteq X_\F$,  there exists string $r\in\L(G,x_0^2)$ from some state $x_0^2\in X_0$ such that $M(st)\!=\!M(r)\!=\!o_1o_2\cdots o_n$ and $\delta(x_0^2,r)\!\notin\! X_\F$. 
Clearly, all states visited by $r$ are in $X_\N$. 
Therefore, based on the construction of $K_G$, there are two finite runs
    \[
    \rho^i=(x_0^i,\epsilon)(x_1^i,o_1)\cdots (x_k^i,o_k)\cdots (x_{k+n}^i,o_{k+n}), i=1,2
    \]
such that $\rho^1 \in \run(st,x_0^1)$, while $\rho^2\in \run(r,x_0^2)$.
Since $n>|X|^2+1$, there must exist two integers $k\leq m_1<m_2\leq k+n$ such that $x_{m_1}^i=x_{m_2}^i$ for both $i=1,2$. Therefore, for each $i=1,2$, we can further define an infinite run
    \begin{flalign}
    \rho_i=&(x_0^i,\epsilon)(x_1^i,o_1)\cdots(x_k^i,o_k)\cdots(x_{m_1}^i,o_{m_1})\notag\\
    &\left((x_{m_1+1}^i,o_{m_1+1})\cdots(x_{m_2}^i,o_{m_2})\right)^\omega\in\run(K_G)\notag
    \end{flalign}
Now consider  $\pi_1=L(\rho_1),\pi_2=L(\rho_2)\in\trace(K_G)$.  We have $\pi_1\models\lozenge\F^{\pi_1}$, $\square(o^{\pi_1}=o^{\pi_2})$, and $\pi_2\models\square\neg\F^{\pi_2}$, which violates $K_G\models\phi_{dia}$. The proof is thus completed.
    \end{proof}

	We show the verification of diagnosability using the following example.
	
\begin{myexm}\upshape
Let us still consider system $G$ shown in Figure \ref{fig-exam-diagpred-G} with Kripke structure $K_G$  shown in Figure~\ref{fig-exam-diagpred-KG}. Here, we further assume that $\Sigma_\F=\{f\}$, i.e., $X_\F=\{2\}$.  One can observe easily that $G$ is diagnosable since one can claim the occurrence of fault immediately after observing symbol $o_2$. 
Now, we show how this is captured by Theorem~\ref{thm-diag} using our framework. 
Taking the negation of $\phi_{dia}$, we have 
\[
\neg \phi_{dia}=\exists\pi_1.\exists\pi_2.\left[\lozenge \F^{\pi_1}\wedge\square(o^{\pi_1}=o^{\pi_2}) \wedge  \square \neg\F^{\pi_2}\right].
\]
To satisfy $\F^{\pi_1}$, 
trace $\pi_1$ must be of form  $\pi_1=\cdots\{2,o_2\}^\omega$, 
while to satisfy $\square \neg\F^{\pi_2}=\neg \lozenge \F^{\pi_2}$, 
trace $\pi_2$ must be of form  $\pi_2=\cdots\{5,o_3\}^\omega$. 
However, this implies that $\square(o^{\pi_1}=o^{\pi_2}) $ cannot be further satisfied. 
Therefore, we have 
$K_G\models \phi_{dia}$. 
\end{myexm}

\subsection{Predictability in HyperLTL}
For the case of predictability, we observe that,  a system is \emph{not} predictable if 
for some string that goes to a boundary state $x_1\in \partial(G)$,  there exists another string that goes to a normal but non-indicator state $x_2\in X_\N\setminus \I(G)$ such that they have the same observation. 
This is because, from the former state $x_1$, a fault event can  occur immediately, while from the latter state $x_2$, some non-fault string can still execute infinitely. 
In the context of traces in Kripke structure $K_G$, the former string can be captured by a trace $\pi_1$ such that $\lozenge \F^{\pi_1}$, while the second string can be captured by a trace $\pi_2$ such that $\neg \lozenge \F^{\pi_2}$. Furthermore, 
the observation equivalence condition is only applied \emph{before} the first occurrence of fault in $\pi_1$. 
Therefore, $\lozenge \F^{\pi_1}$ and the truncated observation equivalence can be captured together by 
$(o^{\pi_1}\!=\!o^{\pi_2})\U \F^{\pi_1}$. 
This suggests that system $G$ is not predictable if 
\[
\exists \pi_1.\exists \pi_2.\left[(o^{\pi_1}=o^{\pi_2})\U \F^{\pi_1}\wedge \neg \lozenge\F^{\pi_2}\right].
\]
Then by taking the negation of the existence of such two traces, we obtain the following main theorem for predictability.

    \begin{tcolorbox} 
        \begin{mythm}[\textbf{HyperLTL for Predictability}]\label{thm-pred}\upshape  
        System $G$ is predictable if and only if $K_G\models\phi_{pre}$,
        where
        \begin{equation}\label{eq-thmpred}
        \phi_{pre}=\forall \pi_1.\forall \pi_2.\left[(o^{\pi_1}=o^{\pi_2})\U \F^{\pi_1}\to\lozenge \F^{\pi_2}\right].
        \end{equation}
        \end{mythm}
    \end{tcolorbox} 
    
    \begin{proof}
    Suppose, for the sake of contraposition, that $K_G\!\not\models\!\phi_{pre}$, i.e., there are two runs
    \[
    \rho^i=(x_0^i,\epsilon)(x_1^i,o_1^i)\cdots\in\run(K_G),i=1,2
    \]
    such that
    (i) $x_j^2\notin X_\F,\forall j\geq0$; and
    (ii) for some $k\geq 1$, we have $x_k^1\in X_\F$ and 
    for any $0\leq j<k$, we have $x_j^1\notin X_\F$ and $o_j^1=o_j^2=:o_j$.
    Based on the construction of $K_G$, for each $i=1,2$, there exist  an initial state $\hat{x}_0^i$ and string $s^i\in\L^\omega(G,\hat{x}_0^i)$ such that $\rho^i\in\run(s^i,\hat{x}_0^i)$.
    Let $s_\partial\in\overline{\{s^1\}}$ be the prefix of $s^1$ such that $\delta(\hat{x}_0^1,s_\partial)\in\partial(G)$. Without loss of generality, we assume that $x_{k\!-\!1}^1\!=\!\delta(\hat{x}_0^1,s_\partial)$; otherwise, we can use $\delta(\hat{x}_0^1,s_\partial)$ to replace $x_{k\!-\!1}^1$ in run $\rho^1$. 
    For any $t\in\overline{\{s_\partial\}}$, there exists $t'\in\overline{\{s^2\}}$ such that $M(t')=M(t)$. However, $\delta(\hat{x}_0^2,t')\notin \I(G)$, since given any $n\in\NN$, we can always choose $t'w'\in\overline{\{s^2\}}$ such that $|w'|> n$ and $\delta(\hat{x}_0^2,t'w')\notin X_\F$. Therefore, we have $\{\delta(\hat{x}_0^1,t),\delta(\hat{x}_0^2,t')\}\subseteq\hat{X}_G(M(t))\not\subseteq\I(G)$. This shows that system $G$ is not predictable.
    
    ($\Leftarrow$)
    We still prove this direction by contraposition. Suppose that $G$ is not predictable, i.e.,
    there exists $s\in\L(G,x_0^1)$ such that $\delta(x_0^1,s)\in\partial(G)$ and $\hat{X}_G(M(t))\not\subseteq\I(G),\forall t\in\overline{\{s\}}$.
    Then there exists a fault event $f\in\Sigma_\F$ such that $\delta(x_0^1,sf)\in X_\F$. 
    We denote $M(sf)=o_1o_2\cdots o_k$ and $x_k^1=\delta(x_0^1,sf)$.
    Based on the construction of $K_G$, we can find a finite run
    \[
    \rho^1=(x_0^1,\epsilon)(x_1^1,o_1)\cdots(x_k^1,o_k) \cdots\in\run(K_G)
    \]
    where $x_k^1\!\in\! X_\F$.
    Choose a prefix $t\!\in\!\overline{\{s\}}$ such that $M(t)\!=\!o_1\cdots o_{k\!-\!1}$.
    Since $\hat{X}_G(M(t))\!\not\subseteq\!\I(G)$, there exist $t'\!\in\!\L(G,x_0^2)$ such that $M(t')\!=\!M(t)\!=\!o_1\cdots o_{k\!-\!1}$ and $\delta(x_0^2,t')\!\notin\!\I(G)$. 
    As such, we can always find an infinite suffix $w'\!\in\!\L^\omega(G,x_0^2)/t'$ such that $\Sigma_\F\!\notin\! t'w'$. We denote $M(w')\!=\!o_k^2o_{k+1}^2\cdots$.
    Based on the construction of $K_G$, we can find a finite run
    \[
    \rho^2=(x_0^2,\epsilon)(x_1^2,o_1^2)\cdots(x_{k\!-\!1}^2,o_{k\!-\!1})(x_k^2,o_k^2)\cdots\in\run(K_G)
    \]
    such that $\rho^2\!\in\!\run(t'w',x_0^2)$. Since $\Sigma_\F\!\notin\! t'w'$, we know that $x_j^2\notin X_\F,\forall j\!\geq\!0$.
    Considering $\pi_1\!=\!L(\rho^1),\pi_2\!=\!L(\rho^2)\!\in\!\trace(K_G)$, we have $(o^{\pi_1}\!=\!o^{\pi_2})\U \F^{\pi_1}\wedge\square \neg\F^{\pi_2}$, which violates $K_G\models\phi_{pre}$. The proof is thus completed.
    \end{proof}
    
\begin{myexm}\upshape
Still, let us consider system $G$ in Figure~\ref{fig-exam-diagpred-G} with Kripke structure $K_G$  shown in Figure~\ref{fig-exam-diagpred-KG}. 
However, this system is not predictable, 
since for string $a$, we cannot alarm the possible occurrence of $f$ in the next step based on observation $o_1$ as the system may also execute string $u_1bu_1$ after which fault will never occur. 
To see how this is captured by Theorem \ref{thm-pred}, we consider the following two traces
\begin{align}
\pi_1= & \{0\}\{1,o_1\}\{2,o_2\}^\omega \in \trace(K_G) \nonumber\\
\pi_2= & \{3\}\{4,o_1\}\{5,o_3\}^\omega \in \trace(\tilde{K}_G) \nonumber 
\end{align}
We have $\lozenge \F^{\pi_1}$ and in fact, before 
$\F^{\pi_1}$ holds, the observation of $\pi_1$ and $\pi_2$ are both $o_1$,
i.e., $(o^{\pi_1}\!=\!o^{\pi_2})\U\F^{\pi_1}$ holds. 
Furthermore,  $\neg \lozenge \F^{\pi_2}\!=\!  \square\neg\F^{\pi_2}$ holds. 
Therefore, for the negation of $\phi_{pre}$, i.e., 
\[
\neg \phi_{pre}=\exists \pi_1.\exists \pi_2.\left[(o^{\pi_1}=o^{\pi_2})\U \F^{\pi_1}\wedge \square \neg \F^{\pi_2}\right], 
\]
we have $K_G\models\neg \phi_{pre}$, which means that   $G$ is not predictable by Theorem \ref{thm-pred}.
\end{myexm}

\section{Detectability in HyperLTL}\label{sec-detect}
Detectability is a property characterizing whether or not the precise state of the system can be determined unambiguously under imperfect observations. Depending on the specific detection requirements, various notions of detectability have been proposed in the literature. In this section, we consider variants of detectability including I-detectability, strong detectability, weak detectability, and delayed detectability, and show how each of   them can be formulated in terms of HyperLTL formula.

First, we review some existing notions of detectability. 
\begin{mydef}[\bf Detectability]\label{def-detect}\upshape
Given system $G$ and observation mask $M:\Sigma\to\O\cup\{\epsilon\}$, we say system $G$ is 
\begin{itemize}
    \item  
	\emph{I-detectable} \cite{shu2012detectability} if the initial-state of the system can always be determined after a finite number of observations, i.e.,
	    \begin{equation}
	    (\exists n\in\NN)(\forall \alpha\in M(\L(G)):|\alpha|\geq n)[|\hat{X}_{G,0}(\alpha)|=1].\notag
	    \end{equation}
	    \item 
	    \emph{strongly detectable} \cite{shu2007detectability} if the current-state of the system can always be determined after a finite number of observations, i.e., 
	    \begin{equation}
	    (\exists n\in\NN)(\forall \alpha\in M(\L(G)):|\alpha|\geq n)[|\hat{X}_G(\alpha)|=1].\notag
	\end{equation}
	    \item 
	    \emph{weakly detectable} \cite{shu2007detectability} if the current-state and the subsequent states of the system can be determined after a finite number of observations for some trajectory of the system, i.e.,
	    \begin{flalign}
	        (\exists n\in\NN)&(\exists \alpha\in M(\L^\omega(G)))\notag\\
	        &(\forall \beta\in\overline{\{\alpha\}}:|\beta|\geq n)[|\hat{X}_G(\beta)|=1].\notag
	    \end{flalign}
	   \item
	   \emph{delayed-detectable} \cite{shu2012delayed} if the precise state of the system at any instant can be determined after some observation delays, i.e.,
	   \begin{equation}
	       (\exists n\!\in\!\NN)(\forall \alpha\beta\!\in\! M(\L(G))\!:\!|\beta|\!\geq\! n)[|\hat{X}_G(\alpha\mid\alpha\beta)|\!=\!1].\notag
	   \end{equation}
	\end{itemize}
	\end{mydef}
	 
Now, we formulate the above variants of detectability using HyperLTL. Based on different types of state-estimates, we present our result in three parts in what follows.
    
\subsection{I-Detectability in HyperLTL}
Still, before expressing detectability in HyperLTL, we define some notation simplifications for HyperLTL formulae.  
For any trace variable $\pi$, we define $X_0^{\pi}$ as the proposition that the trace starts from an initial state in $G$, i.e., $X_0^{\pi}\equiv\bigvee_{x\in X_0}x^{\pi}$. Furthermore, 
for trace variables $\pi_1$ and $\pi_2$, we define  $x^{\pi_1}=x^{\pi_2}$ as the proposition that the  states of $\pi_1$ and $\pi_2$ at the initial instant are equivalent, i.e., 
\[
x^{\pi_1}=x^{\pi_2}\text{ iff }\bigwedge_{x\in X}x^{\pi_1}\leftrightarrow x^{\pi_2}
\] 
We denote by $x^{\pi_1}\!\neq\! x^{\pi_2}$ if $\neg(x^{\pi_1}\!=\!x^{\pi_2})$, which means that $\pi_1[0]$ and $\pi_2[0]$ are not state-equivalent. 

The following theorem states how to formulate I-detectability of $G$ using HyperLTL   for Kripke structure $K_G$. 
\begin{tcolorbox} 
\begin{mythm}[\textbf{HyperLTL for I-Detectability}]\label{thm-id}\upshape  {$\empty$}\\
System $G$ is I-detectable if and only if  $K_G\models \phi_{id}$, 
where
\begin{equation}
\phi_{id}=\forall\pi_1.\forall\pi_2.
\left[
\begin{aligned}
    &[X_0^{\pi_1}\wedge X_0^{\pi_2}\wedge\square(o^{\pi_1}=o^{\pi_2})]\\
    &~~~~~~~\to (x^{\pi_1}=x^{\pi_2})
\end{aligned}
\right].
\end{equation}
 \end{mythm}
\end{tcolorbox} 

Intuitively, the above theorem says that, 
for any two infinite traces in $K_G$ that are initiated from actual initial-states in $G$, 
if they always have the same observation proposition, then they must have the same state proposition initially. Otherwise, there exist two infinite traces starting from two distinct initial-states but having the same observation, which violates the requirement of I-detectability. 
Following this intuition, we present the formal proof of Theorem~\ref{thm-id}.

\begin{proof}($\Rightarrow$)
Suppose, for the sake of contraposition, that  $K_G \!\not\models\! \phi_{id}$, i.e.,  there are two runs 
\[
\rho^i=(x^i_0,\epsilon) (x^i_1,o^i_1)\cdots\in\run(K_G),i=1,2
\]
such that 
(i)  $o^1_j\!=\!o^2_j\!=:\!o_j,\forall j\!\geq\! 1$;   (ii) $x^1_0\!\neq\! x^2_0$; 
and (iii) $x^1_0 \in X_0$ and $x^2_0\in X_0$.  
Based on the construction of $K_G$,   
we can find strings $s^i\!\in\!\L^\omega(G,x_0^i)$ for each $\rho^i$ such that 
$\rho^i\in \run(s^i,x^i_0)$.   
Therefore, for any $n\in \mathbb{N}$, 
one can choose $\alpha=o_1\cdots o_n\in M(\L(G))$ 
such that $\{ x^1_0, x^2_0 \}\subseteq \hat{X}_{G,0}(\alpha)$, i.e., $G$ is not I-detectable.  

($\Leftarrow$)
We still prove this direction by a contrapositive argument. 
Suppose that $G$ is not I-detectable, i.e., 
\[
(\forall n\in\NN)(\exists \alpha\in M(\L(G)):|\alpha|\geq n)[|\hat{X}_{G,0}(\alpha)|>1].
\]
Let us choose $n\!>\!|X|^2\!+\!1$.  
Then we know that there exist two distinct initial states $x^1_0,x^2_0\in X_0$ and  
two strings  $s^i \in \L(G, x^i_0 ),i=1,2$ 
such that 
$M(s^1)\!=\!M(s^2)\!=\!\alpha\!=:\!o_1\cdots o_n$. 
Then for each $i=1,2$,  there is a finite run
\[
\rho^i=(x^i_0,\epsilon)(x^i_{1}, o_1 )(x^i_{2}, o_2)\cdots 
(x^i_{n}, o_n)
\]
such that $\rho^i\in\run(s^i,x_0^i)$.
Since $n>|X|^2+1$, 
there must exist two integers $0\!\leq\!  k_1\!<\!k_2 \!\leq\! m$ such that 
$x^i_{m_1}\!=\!x^i_{m_2}$ for both $i\!=\!1,2$. 
Therefore, for each  $i\!=\!1,2$, we can further define the following infinite run
\begin{align}
\rho_i=
&\;(x^i_0,\epsilon)(x^i_{1}, o_1 )
\cdots(x^i_{m_1}, o_{m_1} ) \nonumber\\
&\;\left((x^i_{m_1+1}, o_{m_1+1} )
\cdots(x^i_{m_2}, o_{m_2} )\right)^\omega \in \run(K_G)\nonumber
\end{align} 
Then by considering $\pi_1\!=\!L(\rho_1), \pi_2\!=\!L(\rho_2)\!\in\! \trace(K_G)$, 
$\phi_{id}$ is violated, which completes the proof. 
\end{proof}

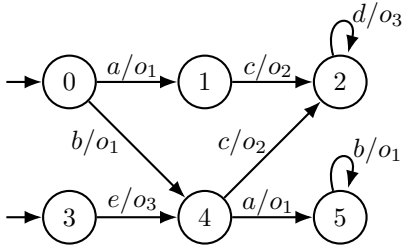
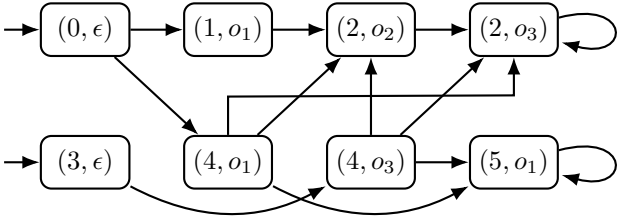
\begin{figure}
		\centering
		\subfigure[System $G$.\label{fig-exam-detect-G}]{
			\centering
			\begin{tikzpicture}[->,>={Latex}, thick, initial text={}, node distance=1.8cm, initial where=left, thick, base node/.style={circle, draw, minimum size=7mm, font=\normalsize}]  
			\node[state, initial, base node, ] (0) {$0$};
			\node[state, base node, ] (1) [right of=0] {$1$};
			\node[state, base node, ] (2) [right of=1] {$2$};
			\node[state, initial, base node, ] (3) [below of=0] {$3$};
			\node[state, base node, ] (4) [right of=3] {$4$};
			\node[state, base node, ] (5) [right of=4] {$5$};
			
			\path[->]
			(0) edge node [xshift=-0.06cm,yshift=0.2cm] {\normalsize $a/o_1$} (1)
			(0) edge node [xshift=-0.55cm, yshift=0.1cm] {\normalsize $b/o_1$} (4)
			(1) edge node [xshift=-0.06cm,yshift=0.2cm] {\normalsize $c/o_2$} (2)
			(2) edge [loop above ] node         [xshift=0.5cm, yshift=-0.2cm]    {\normalsize $d/o_3$} ()
			(5) edge [loop above ] node         [xshift=0.5cm, yshift=-0.2cm]     {\normalsize $b/o_1$} ()
			(3) edge node [xshift=-0.06cm,yshift=0.23cm] {\normalsize $e/o_3$} (4)
			(4) edge node [xshift=-0.06cm,yshift=0.2cm] {\normalsize $a/o_1$} (5)
			(4) edge node [xshift=-0.4cm, yshift=0.1cm] {\normalsize $c/o_2$} (2);
			\end{tikzpicture}
		}
		\\
		\subfigure[\label{fig-exam-detect-KG}Kripke structure $K_G$.]{
			\centering
			\begin{tikzpicture}[->,>={Latex}, thick, initial text={}, node distance=1.9cm, initial where=left, thick, base node/.style={rectangle, rounded corners, draw, minimum size=7mm, font=\normalsize}]   
			\node[initial, state, base node, rectangle, rounded corners, ] (0) {$\;(0,\epsilon)\;$} ;
			\node[state, base node, rectangle, rounded corners, ] [right of=0] (1) {$(1,o_1)$} ;
			\node[state, base node, rectangle, rounded corners, ] [right of=1] (2) {$(2,o_2)$} ;
			\node[state, base node, rectangle, rounded corners, ] [right of=2] (22) {$(2,o_3)$};
			\node[initial, state, base node, rectangle, rounded corners, ] [below=1.4cm] (3)  {$\;(3,\epsilon)\;$};
			\node[state, base node, rectangle, rounded corners, ] [right of=3] (4) {$(4,o_1)$};
			\node[state, base node, rectangle, rounded corners, ] [right of=4] (44) {$(4,o_3)$};
			\node[state, base node, rectangle, rounded corners, ] [right of=44] (5) {$(5,o_1)$};

			\path[->]
			(0) edge node  {} (1)
			    edge node  {} (4)
			(1) edge node  {} (2)
			(2) edge node  {} (22)
			(4) edge node  {} (2)
			(44) edge node  {} (2)
			    edge node  {} (22)
			(44) edge node  {} (5)
			(22)   edge [loop right] node {} ()
			(5)   edge [loop right] node {} ()
			;
			
			\draw[->]
			($(4.north)$) -- ($(4.north)+(0,0.51)$) --
			($(22.south)+(0,-0.51)$) --
			($(22.south)$);
			
			\draw[->]
            (3) to [bend right=28] (44);
            \draw[->]
            (4) to [bend right=28] (5);
			
			\end{tikzpicture}
		}
		\caption{Illustrative example for detectability.}\label{fig-exam-detect}
	\end{figure}

We illustrate Theorem~\ref{thm-id} by the following example. 

\begin{myexm}\upshape
Let us consider system $G$ shown in Figure \ref{fig-exam-detect-G}, where $X\!=\!\{0,1,2,3,4,5\}$, $\Sigma\!=\!\{a,b,c,d,e\}$, $\O\!=\!\{o_1,o_2,o_3\}$   and the observation mask $M\!:\!\Sigma\!\to\!\O$ is defined by: $M(a)\!=\!M(b)\!=\!o_1,M(c)\!=\!o_2$ and $M(d)\!=\!M(e)\!=\!o_3$.
Its Kripke structure $K_G$ is shown in Figure \ref{fig-exam-detect-KG}. 
Clearly, system $G$ is I-detectable since one will first observe symbol $o_1$ if the system starts from initial-state $0$, and   first observe symbol $o_3$ if the system starts from initial-state $3$.  

To see how this is captured by HyperLTL formula $\phi_{id}$ in $K_G$,   
let us consider its negation, i.e., 
\[
\neg \phi_{id}= 
 \exists\pi_1.\exists \pi_2.\left[X_0^{\pi_1}\!\wedge\! X_0^{\pi_2}\!\wedge\!\square(o^{\pi_1}\!=\!o^{\pi_2})\!\wedge\! (x^{\pi_1}\!\neq\! x^{\pi_2})\right].
\]
Clearly, if $x^{\pi_1}\neq x^{\pi_2}$ and $X_0^{\pi_1}\!\wedge\! X_0^{\pi_2}$, then it is not possible to have  
$\square(o^{\pi_1}=o^{\pi_2})$. Therefore, the negation does not hold, i.e., $K_G\models \phi_{id}$, which further implies that $G$ is I-detectable by Theorem~\ref{thm-id}.  
\end{myexm}

\subsection{Strong/Weak Detectability in HyperLTL}
Next, we consider the cases of strong detectability and weak detectability. 
Compared with I-detectability, where two infinite traces need to have the same  state proposition initially, strong detectability requires that two infinite traces need to \emph{converge to} the same state proposition. 
Such a convergence requirement can be captured by the combination  of temporal operators ``\emph{always eventually}" $\lozenge\square$. Recall that, in LTL, 
$\pi \!\models\! \lozenge\square \varphi$, if there exists $i\!\geq\! 0$ 
such that for any $j\!\geq\! i$, we have $\pi[j,\infty]\!\models\! \varphi$. 

Now we present the  following theorem stating  how to formulate strong detectability  using HyperLTL. 

\begin{tcolorbox} 
\begin{mythm}[\textbf{HyperLTL for Strong Detectability}]\label{thm-sd}\upshape  
System $G$ is strongly detectable if and only if  $K_G\models \phi_{sd}$, 
where
\begin{equation}
 \phi_{sd}=\forall\pi_1.\forall\pi_2.\left[\square(o^{\pi_1}=o^{\pi_2})\to\lozenge\square(x^{\pi_1}=x^{\pi_2})\right].
\end{equation}
 \end{mythm}
\end{tcolorbox} 
\begin{proof}($\Rightarrow$)
Suppose, for the sake of contraposition, that $K_G\not\models\phi_{sd}$, i.e., there are two runs
\[
\rho^i=(x^i_0,\epsilon) (x^i_1,o^i_1)\cdots\in\run(K_G),i=1,2
\]
such that (i) $o_j^1\!=\!o_j^2\!=:\!o_j,\forall j\!\geq\! 1$; (ii) $\forall n\!\in\!\N,\exists k\!>\!n\!:\!x_k^1\!\neq\! x_k^2$.
Based on the construction of $K_G$, 
for each $i=1,2$, we can have an initial-state $\hat{x}_0^i\!\in\! X_0$ and a string  $s^i$   such that $\rho^i\!\in\!\run(s^i,\hat{x}_0^i)$. Therefore, for any $n\in\N$, one can choose $\alpha\!=\!o_1\cdots o_k\in M(\L(G))$ such that $\{x_k^1,x_k^2\}\subseteq \hat{X}_G(\alpha)$, i.e., $G$ is not strongly detectable.

($\Leftarrow$) 
Suppose that $K_G\models\phi_{sd}$. Then for every run
\[
\rho=(x_0,\epsilon)(x_1,o_1)(x_2,o_2)\cdots\in\run(K_G)
\]
such that for any other run $\rho'\!\in\!\run(K_G)$ satisfying $\square(o^{L(\rho)}\!=\!o^{L(\rho')})$, we have $\lozenge\square(x^{L(\rho)}\!=\!x^{L(\rho')})$.
Now, for each $j\!\geq\! 1$, we define $\hat{q}_j$ as the current-state estimate of $o_1\cdots o_j$, i.e.,
\begin{equation}\label{eq:pf-c-est}
\hat{q}_j=
\left\{ x'_j\in X
: \!\!\!\!\!\!\!
\begin{array}{cc}
     &  (x'_0,\epsilon)(x'_1,o'_1)\cdots\in \run(K_G)\\
     & 
\text{ s.t. }
o_i'=o_i,\forall 1\leq i\leq j 
\end{array}
\right\} 
\end{equation}
Naturally, we have $\hat{q}_j=\hat{X}_G(o_1\cdots o_j)$.
For infinite sequence $\hat{q}_0\hat{q}_1\hat{q}_2\cdots\in(2^X)^\omega$, we define
\begin{equation}\label{eq:pf-c-est-inf}
\mathbb{X}_{\inf}=
\{
\hat{q}\in 2^X: \forall n\in \NN,\exists j>n\text{ s.t. }\hat{q}_j=\hat{q}
\}
\end{equation}
as the set of state estimates that appear infinite number of times in $\hat{q}_0\hat{q}_1\hat{q}_2\cdots$. Let $n_{\inf}\in\NN$ be the smallest integer such that $\hat{q}_{n_{\inf}}\hat{q}_{n_{\inf}+1}\cdots\in\X^\omega$.
We claim that, for each $\hat{q}\!\in\!\X_{\inf}$, we have $|\hat{q}|\!=\!1$. Otherwise, there would exist $\rho'\in\run(K_G)$ such that $\square(o^{L(\rho)}\!=\!o^{L(\rho')})$ but $\square\lozenge(x^{L(\rho)}\!\neq\! x^{L(\rho')})$.
Now, let us consider any observation string $\alpha\in M(\L(G))$ such that $|\alpha|\!=\!m\!\geq\! n_{\inf}$ and we denote it by $\alpha\!=\!o_1o_2\cdots o_m$. Then we have $|\hat{X}_G(\alpha)|\!=\!|\hat{q}_m|\!=\!1$. Since both of $\rho$ and $\alpha$ are chosen arbitrarily, the following holds
\[
(\forall\alpha\in M(\L(G)):|\alpha|\geq n_{\inf})[|\hat{X}_G(\alpha)|=1]
\]
which proves that $G$ is strongly detectable.
\end{proof}  

The case of weak detectability is similar to the strong counterpart. 
The main difference is that, for strong detectability, we require that 
\emph{for all} traces, we can eventually determines its state, while weak detectability only requires the \emph{existence} of such a trace. 
Therefore,  the HyperLTL condition  $\phi_{wd}$ for weak detectability simply replaces the first universal quantifier $\forall$ in $\phi_{sd}$ by an existential quantifier $\exists$. 
Note that, although Theorem~\ref{thm-wd} seems to be similar to Theorem~\ref{thm-sd}, there is significant difference here: $\forall.\forall.$ in $\phi_{sd}$ does not require quantifier alternation, while $\exists.\forall.$ in $\phi_{wd}$ has one time of quantifier alternation.   

\begin{tcolorbox} 
\begin{mythm}[\textbf{HyperLTL for Weak Detectability}]\label{thm-wd}\upshape  
System $G$ is weakly detectable if and only if  $K_G\models \phi_{wd}$, 
where
\begin{equation}
\phi_{wd}=\exists\pi_1.\forall\pi_2.\left[\square(o^{\pi_1}=o^{\pi_2})\to\lozenge\square(x^{\pi_1}=x^{\pi_2})\right].
\end{equation}
 \end{mythm}
\end{tcolorbox}     
\begin{proof}($\Rightarrow$)
Suppose that $G$ is  weakly detectable, i.e., 
there exists $n\in\NN$ and $\alpha\in M(\L^\omega(G)))$ such that 
\begin{equation}\label{eq:pf-dt1}
(\forall \beta\in\overline{\{\alpha\}}:|\beta|\geq n)[|\hat{X}_G(\beta)|=1]
\end{equation}
Let $x_0\!\in\! X_0$ and 
$s \!\in\! \L^\omega(G,x_0)$ be an infinite string such that 
$M(s)\!=\!\alpha\!=:\!o_1o_2\cdots$.  
Then we can find an infinite run
\[
\rho=(x_0,\epsilon)(x_{\imath_1},o_1)(x_{\imath_2},o_2)\cdots\!\in\!\run(s,x_0)
\]
such that $\rho\!\in\!\run(s,x_0)$. 
We claim that, for $\pi\!=\!L(\rho)$, it holds that 
\begin{equation}\label{eq:pf-dt2}
\forall\pi' \in \trace(K_G).\left[\square(o^{\pi}=o^{\pi'})\to\lozenge\square(x^{\pi}=x^{\pi'})\right]
\end{equation}
Otherwise, it means that there exists a run 
\[\rho'\!=\!(x'_0,\epsilon)(x'_1,o_1)(x'_2,o_2)\cdots\!\in\! \run(K_G)
\]
such that $\square(o^{\pi}\!=\!o^{\pi'})$ but $\square\lozenge(x^{\pi}\!\neq\! x^{\pi'})$. 
Based on the construction of $K_G$, there exists an initial state $\hat{x}'_0\in X_0$ and 
an infinite string $s'\!\in\! \L^\omega(G,\hat{x}_0')$ such that 
$\rho'\!\in\!\run(s',\hat{x}_0')$.
From $\square(o^{\pi}\!=\!o^{\pi'})$, we know $M(s)\!=\!M(s')$.
However, 
$\square\lozenge(x^{\pi}\!\neq\! x^{\pi'})$ means that, for any $n\!\in\! \NN$, there exists 
$t\in\overline{\{s\}},t'\in\overline{\{s'\}}$ such that 
$M(t)\!=\!M(t')\!=:\!\alpha$ but $\delta(x_0,t)\!\neq\!\delta(\hat{x}_0',t')$.  
Then it follows that $\{\delta(x_0,t),\delta(\hat{x}_0',t')\}\subseteq \hat{X}_G( \alpha )$. 
However, this contradicts with Equation~\eqref{eq:pf-dt1}. 
Therefore, we know that Equation~\eqref{eq:pf-dt2} holds, which means $K_G\!\models \!\phi_{wd}$. 

($\Leftarrow$) 
Suppose that $K_G\models \phi_{wd}$, i.e.,    there exists a run 
\[
\rho=(x_0,\epsilon)(x_1,o_1)(x_2,o_2)\cdots\in \run(K_G)
\]
such that for any other run 
$\rho'\!\in\! \run(K_G)$, if $\square(o^{L(\rho)}\!=\!o^{L(\rho')})$, then
we have 
$\lozenge\square(x^{L(\rho)} \!=\!  x^{L(\rho')})$. 
Now, for each $j\geq 1$, we still define $\hat{q}_j$ as the current-state estimate of $o_1\cdots o_j$ as in Equation~\eqref{eq:pf-c-est}, where  we also have $\hat{q}_j\!=\!\hat{X}_G(o_1\cdots o_j)$.
Similarly, for infinite sequence $\hat{q}_0\hat{q}_1\hat{q}_2\cdots\in(2^X)^\omega$, we still define
$\mathbb{X}_{\inf}$ as the set of state estimates that appear infinite number of times in $\hat{q}_0\hat{q}_1\hat{q}_2\cdots$. Let $n_{\inf}\in\NN$ be the smallest integer such that $\hat{q}_{n_{\inf}}\hat{q}_{n_{\inf}+1}\cdots\in\X^\omega$.
We claim that for each $\hat{q}\in\X_{\inf}$, we have $|\hat{q}|=1$. Otherwise, there would exist $\rho'\in\run(K_G)$ such that $\square(o^{L(\rho)}=o^{L(\rho')})$ but $\square\lozenge(x^{L(\rho)}\neq x^{L(\rho')})$.
Now let us choose $\alpha\!=\!o_1o_2\cdots \!\in\! M(\L^\omega(G))$. Then it holds that
\[
(\forall \beta\in \overline{\{\alpha\}}:|\beta|\geq n_{\inf} )
[  |\hat{X}_G(\beta)|=1 ]
\]
which proves that $G$ is weakly detectable. 
\end{proof}     

We illustrate Theorems~\ref{thm-sd} and~\ref{thm-wd} by the following example. 

\begin{myexm}\upshape
Still, let us consider system $G$ shown in Figure \ref{fig-exam-detect-G} with Kripke structure $K_G$  shown in Figure \ref{fig-exam-detect-KG}. 
This system is also strongly detectable since any observation sequence must end  up with either $o_3o_3\cdots$ or $o_1o_1\cdots$: 
the former implies that the system is currently at state $2$ and
the latter implies that the system is currently at state $5$. 
To see how this is captured by formula $\phi_{sd}$, we still consider its negation
\[
\neg \phi_{sd}= \exists\pi_1.\exists\pi_2.\left[\square(o^{\pi_1}=o^{\pi_2})\wedge \square\lozenge(x^{\pi_1}\neq x^{\pi_2})\right].
\]
However, for any traces $\pi_1$ and $\pi_2$ if 
$\square(o^{\pi_1}=o^{\pi_2})$ holds,  then their runs must both loop at state $(2,o_3)$ or $(5,o_1)$, i.e., it is not possible to have
$\square\lozenge(x^{\pi_1}\neq x^{\pi_2})$. 
Therefore, $K_G\models \phi_{sd}$, which means that $G$ is strongly detectable.  Furthermore, since $\phi_{wd}$ is strictly weaker than $\phi_{sd}$, 
we know immediately that $K_G\models \phi_{wd}$, i.e., 
$G$ is also weakly detectable. 
\end{myexm}

\subsection{Delayed Detectability in HyperLTL}\label{subsec-delaydetect}
Finally, we consider the case of delayed detectability, which seems to be more complicated since delayed-state estimate $\hat{X}_G(\alpha \mid \alpha\beta)$  is involved. However, we show that it can still be captured by HyperLTL quite elegantly in a similar fashion as the cases of other notions of detectability.  
To this end, we observe that, system $G$ is not delayed-detectable if there exists an observation $\alpha$ such that $|\hat{X}_G(\alpha \mid \alpha\beta)|\geq 2$ no matter how long the future observation $\beta$ is.  Then by extending observation $\beta$ to the infinite instant, we can obtain two infinite strings in $G$ such that (i) they have the same observation; and (ii) they reach different states at instant $|\alpha|$. 
This key observation leads to the following theorem.

\begin{tcolorbox} 
\begin{mythm}[\textbf{HyperLTL for Delayed Detectability}]\label{thm-dd}\upshape  
System $G$ is delayed-detectable if and only if $K_G\models \phi_{dd}$, 
where
\begin{equation}
\phi_{dd}=\forall\pi_1.\forall\pi_2.\left[
        \square(o^{\pi_1}=o^{\pi_2})\to \square(x^{\pi_1}=x^{\pi_2})
        \right]. 
\end{equation}
 \end{mythm}
\end{tcolorbox} 
\begin{proof}($\Rightarrow$)
Suppose, for the sake of contraposition, that $K_G\!\not\models\!\phi_{dd}$, i.e., there are two runs
\[
\rho^i=(x^i_0,\epsilon) (x^i_1,o^i_1)\cdots(x^i_k,o_k^i)\cdots\in\run(K_G),i=1,2
\]
such that (i) $o_j^1\!=\!o_j^2\!=:\!o_k,\forall k\!\geq\! 1$; (ii) $x_k^1\!\neq\! x_k^2$.
Based on the construction of $K_G$, 
for each $i=1,2$, we can find an initial-state $\hat{x}_0^i$ and  string  $s^i$  such that $\rho^i\in\run(s^i,\hat{x}_0^i)$. 
Now we consider observation string $\alpha\!=\!o_1o_2\cdots o_k$. 
Then for any observation delay $n\!\in\!\NN$, we can choose  $\beta\!=\!o_{k+1}\ldots o_{k+n}$ and we have $\{x_k^1,x_k^2\}\subseteq \hat{X}_G(\alpha\mid\alpha\beta)$, i.e., $G$ is not delayed-detectable.

($\Leftarrow$) We still prove this direction by contraposition. Suppose that $G$ is not delayed-detectable, i.e.
\[
(\forall n\in\NN)(\exists\alpha\beta\in M(\L(G)):|\beta|\geq n)[\hat{X}_G(\alpha\mid\alpha\beta)>1].
\]
Choose $n\!>\!|X|^2\!+\!1$. Then there exist two strings $s_i\!\in\!\L(G),i\!=\!1,2$ such that $M(s^1)\!=\!M(s^2)\!=\!\alpha\beta\!=:\!o_1\cdots o_k o_{k+1}\cdots o_{k+n}$, $k\!=\!|\alpha|$, and $x_{k}^1\neq x_{k}^2$.
For each $i\!=\!1,2$, there is a finite run
\[
\rho^i=(x_0^i,\epsilon)(x_1^i,o_1)\cdots(x_k^i,o_k)\cdots(x_{k+n}^i,o_{k+n})
\]
such that $\rho^i\in\run(s^i,x_0^i)$. 
Since $n\!>\!|X|^2\!+\!1$, there must exist two integers $k\!\leq\! m_1\!<\!m_2\!\leq\! n+k$ such that $x^i_{m_1}\!=\!x^i_{m_2}$ for both $i=1,2$. Therefore, we can further define the following infinite runs
\begin{align}
\rho_i=
&(x_0^i,\epsilon)\cdots(x_k^i, o_k )\cdots(x^i_{m_1}, o_{m_1} )\nonumber\\
&\left((x^i_{m_1+1}, o_{m_1+1} )
\cdots(x^i_{m_2}, o_{m_2} )\right)^\omega\in\run(K_G) \nonumber
\end{align}
Then by considering $\pi_1=L(\rho_1), \pi_2=L(\rho_2)\in \trace(K_G)$, 
$\phi_{dd}$ is violated, which completes the proof. 
\end{proof}

\begin{myexm}\upshape
We consider again the running example $G$ shown in Figure \ref{fig-exam-detect-G} with Kripke structure $K_G$  shown in Figure \ref{fig-exam-detect-KG}. 
This system is, however, not delayed-detectable. 
To see this, we consider observation $\alpha\!=\! o_1\!\in\! M(\mathcal{L}(G))$. 
For any $n\!\geq\! 1$, we can find $\beta\!=\! o_2 (o_3)^{n}$ such that $|\beta|\!>\!n$
but $\hat{X}_G(\alpha\mid \alpha\beta)\!=\!\{1,4\}$ whose cardinality is two.  
This is also captured by the HyperLTL formula $\phi_{dd}$. 
Specifically, we consider the following two infinite traces 
\begin{align}
    \pi_1=&\{0\}\{1,o_1\}\{2,o_2\}\{2,o_3\}^\omega \in \trace(K_G) \nonumber\\
    \pi_2=&\{0\}\{4,o_1\}\{2,o_2\}\{2,o_3\}^\omega\in \trace(K_G)\nonumber
\end{align}
Clearly,  $\square (o^{\pi_1}=o^{\pi_2})$ holds but
$\square (x^{\pi_1}=x^{\pi_2})$ does not hold. Therefore, 
$K_G\not\models \phi_{dd}$, which means that  $G$ is not delayed-detectable by Theorem \ref{thm-dd}.
\end{myexm}

\section{Opacity in HyperLTL}\label{sec-opacity}
Opacity is another important information-flow property describing the privacy and security requirements of the system.  
In this context, it is assumed that there exists an intruder (passive observer) that can also observe the occurrences of events through the observation mask.  
Furthermore, it is assumed that the system has  some ``secret". 
Then opacity captures the confidentiality that  the secret can be revealed to the intruder via the information-flow. 
In the context of DES, the secret of the system is usually modeled as a set of secret states $X_\S\subseteq X$. 
This naturally  partitions the state space as $X=X_\S\dot{\cup} X_\NS$, where $X_\NS=X\backslash X_\S$ is the set of non-secret states. According to what kind of secrets the  system wants to protect, variants of opacity have been proposed in the literature. In this section, we consider the initial-state opacity, infinite-step opacity, and current-state opacity, and formulate all of them in HyperLTL.
	
\begin{mydef}[Opacity]\label{def-opacity}\upshape
Given system $G$, observation mask $M:\Delta\to\O\cup\{\epsilon\}$, and secret states $X_\S\subseteq X$, we say system $G$ is
\begin{itemize}
  \item 
  \emph{initial-state opaque} \cite{saboori2013verification} if the intruder can never know inevitably that the system was initially from a secret state, i.e.,
	    \begin{equation}
	        (\forall\alpha\in M(\L(G)))[\hat{X}_{G,0}(\alpha)\nsubseteq X_\S]. \nonumber
	    \end{equation}
  \item 
  \emph{current-state opaque} \cite{lin2011opacity} if the intruder can never know  inevitably that the system is currently at a secret state, i.e.,
	    \begin{equation}
	        (\forall\alpha\in M(\L(G)))[\hat{X}_G(\alpha)\nsubseteq X_\S].\nonumber
	    \end{equation}
   \item 
   \emph{infinite-step opaque} \cite{saboori2011verification} if the intruder can never know inevitably that the system was at a secret state for any specific instant, i.e.,
	    \begin{equation}
	        (\forall\alpha\beta\in M(\L(G)))[\hat{X}_{G,0}(\alpha\mid\alpha\beta)\nsubseteq X_\S].\nonumber
	    \end{equation}
\end{itemize}
\end{mydef}
	
For the sake of simplicity, we denote by $\S^\pi$  and $\NS^\pi$ the propositions that the trace is at a secret state and a non-secret state, respectively, i.e.,  
  $\S^\pi\equiv \bigvee_{x\in X_\S}x^{\pi}$ and         $\NS^\pi\equiv \bigvee_{x\in X_\NS}x^{\pi}$. 
Now, we show how these three variants of opacity can be formulated in terms of HyperLTL.  
	
\subsection{Initial-State Opacity in HyperLTL}
Essentially, initial-state opacity requires that, for any string, if it is initiated from a secret state, then there must exist another string such that (i) it is initiated from a non-secret state; and (ii) the two strings have the same observation. 
This requirement can be captured easily by HyperLTL formula based on the Kripke structure $K_G$ as follows. 
 
	\begin{tcolorbox} 
        \begin{mythm}[\textbf{HyperLTL for Initial-State Opacity}]\label{thm-iso}\upshape  
        System $G$ is initial-state opaque if and only if $K_G\models \phi_{iso}$, 
        where
        \begin{equation}
        \phi_{iso}=\forall\pi_1.\exists\pi_2.\left[
        \begin{aligned}
            &[X_0^{\pi_1}\wedge X_0^{\pi_2}\wedge\S^{\pi_1}]\to\\
            &[\square(o^{\pi_1}=o^{\pi_2})\wedge \NS^{\pi_2}]
        \end{aligned}
        \right].
        \end{equation}
        \end{mythm}
    \end{tcolorbox} 
    \begin{proof}
    ($\Rightarrow$) Suppose, for the sake of contraposition, that $K_G\!\not\models\!\phi_{iso}$. 
    Then there exists $x_0\in X_0\cap X_\S$ and a run
    \[
    \rho=(x_0,\epsilon)(x_1,o_1)(x_2,o_2)\cdots\in\run(K_G)
    \]
    such that for any other run
    \[
    \rho'=(x'_0,\epsilon)(x'_1,o_1)(x'_2,o_2)\cdots\in\run(K_G)
    \]
    satisfying $X_0^{L(\rho')}\wedge\square(o^{L(\rho)}\!=\!o^{L(\rho')})$, we have   $x'_0\!\in\! X_\S$.
    Now, for each $j\!\geq\!1$, we define $\hat{q}_j^0$ as the initial-state estimate of $o_1\cdots o_j$, i.e.,
    \begin{equation}\label{eq-qk0hat}
    \hat{q}_j^0=
    \left\{ x'_0\in X
    \!: \!\!\!\!\!\!\!\!\!
    \begin{array}{cc}
        &  (x'_0,\epsilon)(x'_1,o'_1)\cdots\in \run(K_G)\\
        & 
    \text{ s.t. }
    (x_0'\!\in\! X_0)\wedge (o'_i\!=\!o_i),\forall 1\!\leq\! i\!\leq\! j 
    \end{array}
    \!\!\right\} 
    \end{equation}
    By construction, we have
        (i) $\hat{q}_j^0\!=\!\hat{X}_{G,0}(o_1\cdots o_j)$; and
        (ii) $\hat{q}_{j+1}^0\!\subseteq\! \hat{q}_j^0$.
    For infinite sequence $\hat{q}_0^0\hat{q}_1^0\hat{q}_2^0\cdots\in(2^X)^\omega$, we define
    \[
    \mathbb{X}_{\inf}^0=
    \{
    \hat{q}^0\in 2^X: \forall n\in \NN,\exists j>n\text{ s.t. }\hat{q}_j^0=\hat{q}^0
    \}
    \]
    as the set of state estimates that appear infinite number of times in $\hat{q}_0^0\hat{q}_1^0\hat{q}_2^0\cdots$. Let $n_{\inf}\!\in\!\NN$ be the smallest integer such that $\hat{q}_{n_{\inf}}^0\hat{q}_{n_{\inf}+1}^0\cdots\!\in\!(\X_{\inf}^0)^\omega$.
    We claim that, for each $\hat{q}^0\!\in\!\X_{\inf}^0$, we have $\hat{q}^0\subseteq X_\S$. Otherwise, there would exist $\rho'\!\in\!\run(K_G)$ such that $\NS^{L(\rho')}\!\wedge\!\square(o^{L(\rho')}\!=\!o^{L(\rho)})$ holds.
    Now, let us consider  observation $\alpha\!=\!o_1o_2\ldots o_{n_{\inf}}\!\in\! M(\L(G))$. We have $\hat{X}_{G,0}(\alpha)\!=\!\hat{q}_{n_{\inf}}^0\subseteq X_\S$, i.e., $G$ is not initial-state opaque.
    
    ($\Leftarrow$)
    We still prove this direction by contraposition.
    Suppose that system $G$ is not initial-state opaque, which means that there is an initial state $x_0\!\in\! X_\S$ and an observation string $\alpha\!=:\!o_1o_2\cdots o_n\!\in\! M(\L(G,x_0))$ such that $\hat{X}_{G,0}(\alpha)\subseteq X_\S$. 
    Since $G$ is live, one can easily obtain an infinite observation string $\alpha'\!=\!o_1o_2\cdots o_n\cdots\!\in\! M(\L^\omega(G))$. Based on the construction of $K_G$, there is naturally a run
    \[
    \rho=(x_0,\epsilon)(x_1,o_1)(x_2,o_2)\cdots\in\run(K_G)
    \]
    Moreover, we have $\hat{q}_j^0\subseteq \hat{q}_n^0=\hat{X}_G(\alpha)\subseteq X_\S,\forall j\!\geq\! n$, where $\hat{q}_j^0$ is defined in Equation \eqref{eq-qk0hat}. Then we claim that, for $\pi=L(\rho)$, it holds that
    \begin{equation}\label{eq-notiso}
        \forall\pi'\in\trace(K_G).\left[X_0^{\pi'}\wedge\square(o^{\pi}=o^{\pi'})\to \S^{\pi'}\right]
    \end{equation}
    Otherwise, it means that there would be an infinite run $\rho'\!=\!(x'_0,\epsilon)(x'_1,o_1)\cdots\in\run(K_G)$ where $x'_0\!\in\! X_0\!\cap\! X_\NS$. Then, we have $\{x_0,x'_0\}\subseteq\hat{q}_j^0\not\subseteq X_\S,\forall j\geq1$. This immediately contradicts with $\hat{q}_j^0\subseteq X_\S,\forall j\geq n$. Therefore, we know that Equation \eqref{eq-notiso} holds, which shows that $K_G\not\models\phi_{iso}$. The proof is thus completed.
    \end{proof}
    
\subsection{Current-State Opacity in HyperLTL}
When it refers to current-state opacity, however, the following difficulty arises if we want to write down HyperLTL formula that is checked on Kripke structure $K_G$. 
Specifically, current-state opacity requires that for any \emph{finite string} that ends up with a secret state, there exists another finite string ending up with a non-secret state such that they have the same observation. 
However, the semantics of HyperLTL are defined over \emph{infinite traces}. 
To capture the above requirement using HyperLTL, we need some mechanism to \emph{indicate} that two infinite traces are at secret and non-secret states, respectively, at the \emph{same instant}.  Furthermore, the observation equivalence requirement should be only applied up to that indicator instant, not for the entire infinite horizon.

	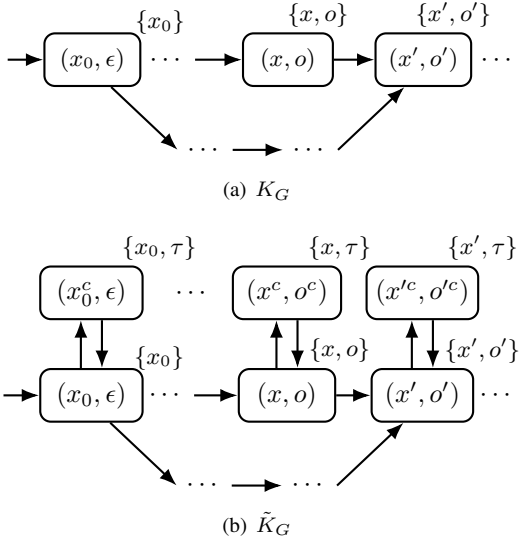
\begin{figure}
		\centering
		\subfigure[$K_G$]{
			\centering
			\begin{tikzpicture}[->,>={Latex}, thick, initial text={}, node distance=1.8cm, initial where=left, thick, base node/.style={rectangle, rounded corners, draw, minimum size=7mm, font=\small}]  
			\node[initial, state, base node,  ] (0) {$\;(x_0,\epsilon)\;$} ;
			\node[ ] at ($(0.center)+(1.5cm,-1.2cm)$) (2)  {$\cdots$};
			\node[ ] at ($(0.center)+(1cm,0)$) (3)  {$\cdots$};
			\node[state, base node,  ] at ($(3.center)+(1.6cm,0)$) (4)  {$~(x,o)~$};
			\node[state, base node,  ]  at ($(4.center)+(1.8cm,0)$) (5)  {$\,(x',o')\,$};
			\node[ ] at ($(5.center)+(1cm,0)$) (6)  {$\cdots$};
			
			\node[ ] at ($(5.center)+(-1.5cm,-1.2cm)$) (8)  {$\cdots$};
			
			\node[font=\small ] at ($(0.center)+(0.9cm,0.5cm)$) (0label)  {$\{x_0\}$};
			\node[font=\small ] at ($(4.center)+(0.4cm,0.6cm)$) (xlabel)  {$\{x,o\}$};
			\node[font=\small ] at ($(5.center)+(0.4cm,0.6cm)$) (xxlabel)  {$\{x',o'\}$};
			
			\path[->]
			(3) edge node  {} (4)
			(4) edge node  {} (5)
			(2) edge node  {} (8)
			;
			
			\draw[->]
			($(0.south)+(0.25cm,0)$) -- ($(2.west)$);
			\draw[->]
			($(8.east)$) -- ($(5.south)+(-0.25cm,0)$);

			\end{tikzpicture}
		}
		\\
		\subfigure[$\tilde{K}_G$]{
			\centering
			\begin{tikzpicture}[->,>={Latex}, thick, initial text={}, node distance=1.8cm, initial where=left, thick, base node/.style={rectangle, rounded corners, draw, minimum size=7mm, font=\normalsize}]  
			\node[initial, state, base node,  ] (0) {$\;(x_0,\epsilon)\;$} ;
			\node[ ] at ($(0.center)+(1.5cm,-1.2cm)$) (2)  {$\cdots$};
			\node[ ] at ($(0.center)+(1cm,0)$) (3)  {$\cdots$};
			\node[state, base node,  ] at ($(3.center)+(1.6cm,0)$) (4)  {$~(x,o)~$};
			\node[state, base node,  ]  at ($(4.center)+(1.8cm,0)$) (5)  {$\,(x',o')\,$};
			\node[ ] at ($(5.center)+(1cm,0)$) (6)  {$\cdots$};
			
			\node[ ] at ($(5.center)+(-1.5cm,-1.2cm)$) (8)  {$\cdots$};
			
			\node[font=\small ] at ($(0.center)+(0.9cm,0.5cm)$) (0label)  {$\{x_0\}$};
			\node[font=\small ] at ($(4.center)+(0.7cm,0.6cm)$) (xlabel)  {$\{x,o\}$};
			\node[font=\small ] at ($(5.center)+(0.8cm,0.6cm)$) (xxlabel)  {$\{x',o'\}$};
			
			\node[state, base node, ] [above=1cm] (0copy)  {$\;(x_0^c,\epsilon)\;$};
			\node[state, base node, ] at ($(4.north)+(0,1cm)$) (xcopy)  {$\,(x^c,o^c)\,$};
			\node[state, base node, ] at ($(5.north)+(0,1cm)$) (xxcopy)  {$(x'^c,o'^c)$};
			\node[ ] at ($(0copy.center)+(1.35cm,0)$) (9)  {$\cdots$};
			
			\node[font=\small ] at ($(0copy.center)+(0.9cm,0.6cm)$) (0copylabel)  {$\{x_0,\tau\}$};
			\node[font=\small ] at ($(xcopy.center)+(0.7cm,0.6cm)$) (xcopylabel)  {$\{x,\tau\}$};
			\node[font=\small ] at ($(xxcopy.center)+(0.8cm,0.6cm)$) (xxcopylabel)  {$\{x',\tau\}$};
			
			\path[->]
			(3) edge node  {} (4)
			(4) edge node  {} (5)
			(2) edge node  {} (8)
			;
			
			\draw[->]
			($(0.south)+(0.25cm,0)$) -- ($(2.west)$);
			\draw[->]
			($(8.east)$) -- ($(5.south)+(-0.25cm,0)$);
			
			\draw[->]
			($(0.north)+(-0.14cm,0)$) -- ($(0copy.south)+(-0.14cm,0)$);
			\draw[->]
			($(0copy.south)+(0.14cm,0)$) -- ($(0.north)+(0.14cm,0)$);
			\draw[->]
			($(4.north)+(-0.14cm,0)$) -- ($(xcopy.south)+(-0.14cm,0)$);
			\draw[->]
			($(xcopy.south)+(0.14cm,0)$) -- ($(4.north)+(0.14cm,0)$);
			\draw[->]
			($(5.north)+(-0.14cm,0)$) -- ($(xxcopy.south)+(-0.14cm,0)$);
			\draw[->]
			($(xxcopy.south)+(0.14cm,0)$) -- ($(5.north)+(0.14cm,0)$);
			\end{tikzpicture}
		}
		\caption{Conceptual illustration of how to construct $\tilde{K}_G$ from $K_G$.}\label{fig-transition23}
	\end{figure}
In order to bridge the  gap between the finite requirement in current-state opacity and the infinite semantics of HyperLTL, we modifies the Kripke structure $K_G$ by allowing the process to stop at any finite instant. 
  
\begin{mydef}[\bf Modified  Kripke Structure]\upshape
Given Kripke structure $K_G=(Q,Q_0,\Delta,\AP,L)$ for DES $G$, 
we defined the \emph{modified Kripke structure}
    \[
    \tilde{K}_G=(\tilde{Q},\tilde{Q}_0,\tilde{\Delta},\widetilde{\AP},\tilde{L})
    \]
    where
    \begin{itemize}
    \item 
    $\tilde{Q}=Q\cup Q^c$ is the set of states, where $ Q^c=\{(x^c,o^c):(x,o)\in Q\}$ is simply a copy of the original state set $Q$; 
    \item 
    $\tilde{Q}_0=Q_0$ is the set of initial states;
    \item 
    $\tilde{\Delta}\subseteq \tilde{Q}\times \tilde{Q}$ is the transition function defined as follows:
    \begin{itemize}
        \item   
        for any $q,q'\in Q:\langle q,q' \rangle\in\tilde{\Delta}$, we have
        $ \langle q,q' \rangle\in\tilde{\Delta} $; 
        \item   
        for any $q=(x,o)\in Q$, we have
        \begin{align}
            \langle (x, o) , (x^c,o^c) \rangle,\langle (x^c,o^c), (x, o)     \rangle &\in \tilde{\Delta}  
        \end{align}
    \end{itemize} 
    \item 
        $\widetilde{\AP}=X\cup\O\cup\{\tau\}$ is the set of atomic propositions, where $\tau$ is a new symbol;
    \item 
        $\tilde{L}:\tilde{Q}\to 2^{\widetilde{\AP}}$ is the labeling function defined by: 
    \begin{itemize}
      \item 
      for any $q\in Q$, we have  $\tilde{L}(q)=L(q)$;  
	\item 
	  for any $q^c=(x^c,o^c)\in Q^c$, we have $\tilde{L}(q^c)=\{ x,\tau \}$.  
\end{itemize}
  \end{itemize}	
	\end{mydef}

In Figure~\ref{fig-transition23}, we show conceptually how we modify Kripke structure $K_G$	to obtain  $\tilde{K}_G$. Specifically, we simply add a new copy state $(x^c,o^c)$ for each state $(x,o)$ in $K_G$. In addition to the original transitions in $K_G$, each state  $(x,o)$ and its copy state $(x^c,o^c)$ form a loop. 
Furthermore, for each copy state, we assign it a new atomic proposition $\tau$. 
Intuitively, $\tau$ will be used as an \emph{indicator} to locate the specific instant of our interest for checking secret status. 

To formally see this, consider an arbitrary run in $K_G$
\[
\rho=(x_0,\epsilon)(x_1,o_1)\cdots\in \run(K_G)
\]
Then for any instant  $k\geq 0$ of our interest for checking the current secret status, based on the construction of $\tilde{K}_G$, there exists the following run in $\tilde{K}_G$
\[
\tilde{\rho}\!=\!(x_0,\epsilon)(x_1,o_1)\!\cdots\!\underbrace{(x_k,o_k)(x_k^c,o_k^c)(x_k,o_k)}_{\text{loop to copy state at instant $k$}}(x_{k+1},o_{k+1})\!\cdots
\]
whose trace is given by
\[
\tilde{L}(\tilde{\rho})\!=\!\{x_0\}\{x_1,o_1\}\!\cdots\!\{x_k,o_k\}\{x_k,\tau\}\{x_k,o_k\}\!\cdots
\]
As such, with the help of atomic proposition $\tau$, we can easily locate $x_k$ for which we what to check whether or not $x_k\in X_\S$, and truncate the infinite sequence after $\tau$. 

Note that, in the above infinite trace, we only want to loop at the copy state \emph{once} at the specific instant of interest. 
However, temporal operator $\lozenge \tau$ is not sufficient to express this since $\tau$ may occur multiple times. To this end, we define operator $\lozenge_1$ as ``eventually and only once" as follows:
\begin{equation}
\lozenge_{1} \tau  \equiv \lozenge\tau\wedge\square(\tau\to\bigcirc\square\neg\tau),  
\end{equation}
i.e., $\tau$ will eventually occur and once it occurs, it will never occur in the future. 
%
    
Now, we formulate current-state opacity as a HyperLTL formula checked on $\tilde{K}_G$ as follows.
	
\begin{tcolorbox} 
\begin{mythm}[\textbf{HyperLTL for Current-State Opacity}]\label{thm-cso}\upshape  
System $G$ is current-state opaque if and only if $\tilde{K}_G\models \phi_{cso}$, 
where
\begin{align}\label{eq:cso-thm}
&\phi_{cso}= \\
&\forall\pi_1.\exists\pi_2.\left[\!\!\! \!\!\!\! \!\! 
        \begin{array}{l l}
             &  \left[  
        \lozenge_1 \tau^{\pi_1} \wedge\square (\tau^{\pi_1}\!\to\! \S^{\pi_1} ) 
        \right]  \to\\
       &  
        \left[
        (o^{\pi_1}\!=\!o^{\pi_2})\U \tau^{\pi_1}
        \wedge 
        \square(\tau^{\pi_1}\!\to\! (\tau^{\pi_2}\!\wedge\! \NS^{\pi_2}) ) 
        \right] 
        \end{array} \!\! 
        \right].  \nonumber
\end{align}
\end{mythm}
\end{tcolorbox} 
  
Formula $\phi_{cso}$ in the above theorem is explained as follows. For trace $\pi_1$, which is quantified by the universal quantifier, we require that it visits a copy state only once, i.e., $\lozenge_1 \tau^{\pi_1}$, and the copy state it visits is a secret state, i.e., $\square (\tau^{\pi_1}\!\to\! \S^{\pi_1} ) $.  Then for such an arbitrary $\pi_1$, we require the existence of trace $\pi_2$ such that 
(i) it has the same observation with $\pi_1$ until the stopping instant, i.e., $(o^{\pi_1}\!=\!o^{\pi_2})\U \tau^{\pi_1}$; 
and 
(ii) when it stops, it is at a non-secret copy state, i.e., 
$\square(\tau^{\pi_1}\!\to\! (\tau^{\pi_2}\!\wedge\! \NS^{\pi_2}) )$. 
With this intuition in mind, we present the formal proof of Theorem~\ref{thm-cso}. 
    \begin{proof}
    ($\Rightarrow$)
    Suppose, for the sake of contraposition, that $\tilde{K}_G\not\models\phi_{cso}$, i.e., there exists a run
    \[
    \rho=(x_0,\epsilon)(x_1,o_1)\cdots(x_n,o_n)(x^c_n,o^c_n)\cdots\in\run(\tilde{K}_G)
    \]
    such that 
     $\tilde{L}((x^c_n,o^c_n))\!=\!\{x_n,\tau\}$ and $x_n\!\in\! X_\S$, 
     i.e., $\lozenge_1 \tau^{\pi} \wedge\square (\tau^{\pi}\!\to\! \S^{\pi} ) $ holds for trace $\pi=\tilde{L}(\rho)$, 
     and
     for any other run  
    \[
    \rho'=(x'_0,\epsilon)(x'_1,o_1)\cdots(x'_n,o_n)(x'^c_n,o^c_n)\cdots\in\run(\tilde{K}_G), 
    \]
    i.e., 
    $[(o^{\pi}\!=\!o^{\pi'})\U\tau^{\pi}]\wedge[\square(\tau^{\pi}\!\to\!\tau^{\pi'})]$ holds for trace $\pi'\!=\!\tilde{L}(\rho')$, 
    we have
    $x'_n\in X_\S$.  Now, for each $j\!\geq\! 1$, we define $\hat{q}_j$ as the current-state estimate of $o_1\cdots o_j$, i.e.,
    \begin{equation}\label{eq-qkhat}
    \hat{q}_j=
    \left\{ x'_j\in X
    : \!\!\!\!\!\!\!
    \begin{array}{cc}
        &  (x'_0,\epsilon)(x'_1,o'_1)\cdots\in \run(\tilde{K}_G)\\
        & 
    \text{ s.t. }
    o'_i=o_i,\forall 1\leq i\leq j
    \end{array}
    \right\} 
    \end{equation}
    By construction, we have $\hat{q}_j\!=\!\hat{X}_G(o_1\cdots o_j)$. We claim that $\hat{q}_n\subseteq X_\S$. Otherwise, there would exists $\rho'\!\in\!\run(\tilde{K}_G)$ such that for $\pi'\!=\!\tilde{L}(\rho')$, $(o^{\pi'}\!=\!o^{\pi})\U\tau^{\pi}\wedge\square(\tau^{\pi}\to\tau^{\pi'})$ but $\square(\tau^{\pi}\to\NS^{\pi'})$
    Let us consider  observation string $\alpha\!=\!o_1\cdots o_n\!\in\! M(\L(G))$. It holds that $\hat{X}_G(\alpha)\subseteq X_\S$, which makes $G$ not current-state opaque.
    
    ($\Leftarrow$)
    We still prove this direction by contraposition. Suppose that $G$ is not current-state opaque, which means that there exists an observation string $\alpha\!=:\!o_1\cdots o_n\!\in\! M(\L(G))$ such that $\hat{X}_G(\alpha)\subseteq X_\S$.  
    Based on the construction of $\tilde{K}_G$, there exists a run
    \[
    \rho\!=\!(x_0,\epsilon)(x_1,o_1)\cdots(x_n,o_n)(x^c_n,o^c_n)\cdots\in\run(\tilde{K}_G)
    \]
    such that $\lozenge_1 \tau^{\pi}\wedge\square(\tau^{\pi}\to\S^{\pi})$ for $\pi\!=\!\tilde{L}(\rho)$,
    and $\hat{q}_n\!=\!\hat{X}_G(\alpha)\!\subseteq\! X_\S$, where $\hat{q}_n$ is defined in Equation \eqref{eq-qkhat}. Then it holds that
    \[
    \forall\pi'\in\trace(\tilde{K}_G).
    \left[
    \begin{aligned}
    &[(o^{\pi}=o^{\pi'})\U\tau^{\pi}\wedge\square(\tau^{\pi}\to\tau^{\pi'})]\\
    &~~\to[\square(\tau^{\pi}\to(\tau^{\pi'}\wedge\S^{\pi'}))]
    \end{aligned}
    \right].
    \]
    This shows that $\tilde{K}_G\not\models\phi_{cso}$ and completes the proof.
    \end{proof}
	\begin{figure}
		\centering
		\subfigure[System $G$\label{fig-exam-opacity-G}]{
			\centering
			\begin{tikzpicture}[->,>={Latex}, thick, initial text={}, node distance=1.8cm, initial where=left, thick, base node/.style={circle, draw, minimum size=7mm, font=\normalsize}]   
			\node[state, initial, base node, fill=red!70] (0) {$0$};
			\node[state, base node, ] (1) [right of=0] {$1$};
			\node[state, base node, ] (2) [right of=1] {$2$};
			\node[state, initial, base node, ] (3) [below of=0] {$3$};
			\node[state, base node, fill=red!70] (4) [right of=3] {$4$};
			\node[state, base node, ] (5) [right of=4] {$5$};
			
			\path[->]
			(0) edge node [xshift=-0.06cm,yshift=0.2cm] {\normalsize $a/o_1$} (1)
			(1) edge node [xshift=-0.06cm,yshift=0.2cm] {\normalsize $c/o_2$} (2)
			(2) edge [loop above ] node         [xshift=0.5cm, yshift=-0.2cm]   {\normalsize $d/o_3$} ()
			(5) edge [loop above ] node         [xshift=0.5cm, yshift=-0.2cm]   {\normalsize $d/o_3$} ()
			(3) edge node [xshift=-0.06cm,yshift=0.2cm] {\normalsize $b/o_1$} (4)
			(4) edge node [xshift=-0.06cm,yshift=0.2cm] {\normalsize $e/o_4$} (5)
			(4) edge node [xshift=-0.4cm, yshift=0.1cm] {\normalsize $c/o_2$} (2);
			\end{tikzpicture}
		}
		\\
		\subfigure[Kripke structure $K_G$\label{fig-exam-opacity-KG}]{
			\centering
			\begin{tikzpicture}[->,>={Latex}, thick, initial text={}, node distance=1.9cm, initial where=left, thick, base node/.style={rectangle, rounded corners, draw, minimum size=7mm, font=\normalsize}]  
			\node[initial, state, base node, rectangle, rounded corners, fill=red!70] (0) {$~(0,\epsilon)~$} ;
			\node[state, base node, rectangle, rounded corners, ] [right of=0] (1) {$\,(1,o_1)\,$} ;
			\node[state, base node, rectangle, rounded corners, ] [right of=1] (2) {$\,(2,o_2)\,$};
			\node[state, base node, rectangle, rounded corners, ] [right of=2] (22) {$\,(2,o_3)\,$};
			\node[initial, state, base node, rectangle, rounded corners, ] [below=1.3cm] (3)  {$~(3,\epsilon)~$};
			\node[state, base node, rectangle, rounded corners, fill=red!70] [right of=3] (4) {$\,(4,o_1)\,$};
			\node[state, base node, rectangle, rounded corners, ] [right of=4] (5) {$\,(5,o_4)\,$};
			\node[state, base node, rectangle, rounded corners, ] [right of=5] (55) {$\,(5,o_3)\,$};
			
			\node[state, base node, rectangle, rounded corners, fill=red!70] [above=1cm] (0copy)  {$\,(0^c,\epsilon)\,$};
			\node[state, base node, rectangle, rounded corners] [right of=0copy] (1copy) {$(1^c,o_1^c)$};
			\node[state, base node, rectangle, rounded corners] [right of=1copy] (2copy) {$(2^c,o_2^c)$};
			\node[state, base node, rectangle, rounded corners] [right of=2copy] (22copy) {$(2^c,o_3^c)$};
			
			\node[state, base node, rectangle, rounded corners, ] at ($(3.south)+(0,-1cm)$) (3copy)  {$\,(3^c,\epsilon)\,$};
			\node[state, base node, rectangle, rounded corners, fill=red!70] [right of=3copy] (4copy) {$(4^c,o_1^c)$};
			\node[state, base node, rectangle, rounded corners] [right of=4copy] (5copy) {$(5^c,o_4^c)$};
			\node[state, base node, rectangle, rounded corners] [right of=5copy] (55copy) {$(5^c,o_3^c)$};
			
			\node[] [xshift=0.68cm, yshift=0.5cm] (0label) {\footnotesize$\{0\}$};
			\node[] [right of=0label] (1label) {\footnotesize$\{1,o_1\}$};
			\node[] [right of=1label] (2label) {\footnotesize$\{2,o_2\}$};
			\node[] [right of=2label] (22label) {\footnotesize$\{2,o_3\}$};
			\node[] [xshift=0.68cm, yshift=-2.2cm] (3label) {\footnotesize$\{3\}$};
			\node[] [right of=3label] (4label) {\footnotesize$\{4,o_1\}$};
			\node[] [right of=4label] (5label) {\footnotesize$\{5,o_4\}$};
			\node[] [right of=5label] (55label) {\footnotesize$\{5,o_3\}$};
			
			\node[] [xshift=0.68cm, yshift=1.9cm] (0copylabel) {\footnotesize$\{0,\tau\}$};
			\node[] [right of=0copylabel] (1copylabel) {\footnotesize$\{1,\tau\}$};
			\node[] [right of=1copylabel] (2copylabel) {\footnotesize$\{2,\tau\}$};
			\node[] [right of=2copylabel] (22copylabel) {\footnotesize$\{2,\tau\}$};
			\node[] [xshift=0.68cm, yshift=-3.6cm] (3copylabel) {\footnotesize$\{3,\tau\}$};
			\node[] [right of=3copylabel] (4copylabel) {\footnotesize$\{4,\tau\}$};
			\node[] [right of=4copylabel] (5copylabel) {\footnotesize$\{5,\tau\}$};
			\node[] [right of=5copylabel] (55copylabel) {\footnotesize$\{5,\tau\}$};
			
			\path[->]
			(0) edge node  {} (1)
			(1) edge node  {} (2)
			(2) edge node  {} (22)
			(22)   edge [loop right] node {} ()
			(3) edge node  {} (4)
			(4) edge node  {} (5)
			(5) edge node  {} (55)
			(55)   edge [loop right] node {} ()
			(4) edge node  {} (2)
			;
			
			\draw[->]
			($(4.north)$) -- ($(4.north)+(0,0.431)$) --
			($(22.south)+(0,-0.431)$) --
			($(22.south)$);
			
			\draw[->]
			($(0.north)+(-0.13cm,0)$) -- ($(0copy.south)+(-0.13cm,0)$);
			\draw[->]
			($(0copy.south)+(0.13cm,0)$) -- ($(0.north)+(0.13cm,0)$);
			\draw[->]
			($(1.north)+(-0.13cm,0)$) -- ($(1copy.south)+(-0.13cm,0)$);
			\draw[->]
			($(1copy.south)+(0.13cm,0)$) -- ($(1.north)+(0.13cm,0)$);
			\draw[->]
			($(2.north)+(-0.13cm,0)$) -- ($(2copy.south)+(-0.13cm,0)$);
			\draw[->]
			($(2copy.south)+(0.13cm,0)$) -- ($(2.north)+(0.13cm,0)$);
			\draw[->]
			($(22.north)+(-0.13cm,0)$) -- ($(22copy.south)+(-0.13cm,0)$);
			\draw[->]
			($(22copy.south)+(0.13cm,0)$) -- ($(22.north)+(0.13cm,0)$);
			\draw[->]
			($(3.south)+(-0.13cm,0)$) -- ($(3copy.north)+(-0.13cm,0)$);
			\draw[->]
			($(3copy.north)+(0.13cm,0)$) -- ($(3.south)+(0.13cm,0)$);
			\draw[->]
			($(4.south)+(-0.13cm,0)$) -- ($(4copy.north)+(-0.13cm,0)$);
			\draw[->]
			($(4copy.north)+(0.13cm,0)$) -- ($(4.south)+(0.13cm,0)$);
			\draw[->]
			($(5.south)+(-0.13cm,0)$) -- ($(5copy.north)+(-0.13cm,0)$);
			\draw[->]
			($(5copy.north)+(0.13cm,0)$) -- ($(5.south)+(0.13cm,0)$);
			\draw[->]
			($(55.south)+(-0.13cm,0)$) -- ($(55copy.north)+(-0.13cm,0)$);
			\draw[->]
			($(55copy.north)+(0.13cm,0)$) -- ($(55.south)+(0.13cm,0)$);
			\end{tikzpicture}
		}
		\caption{Example for opacity.}
	\end{figure}
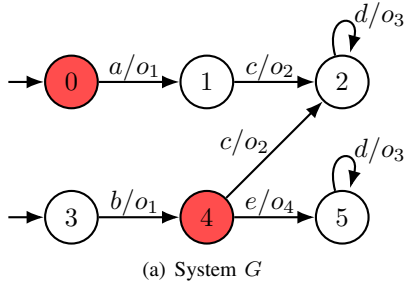
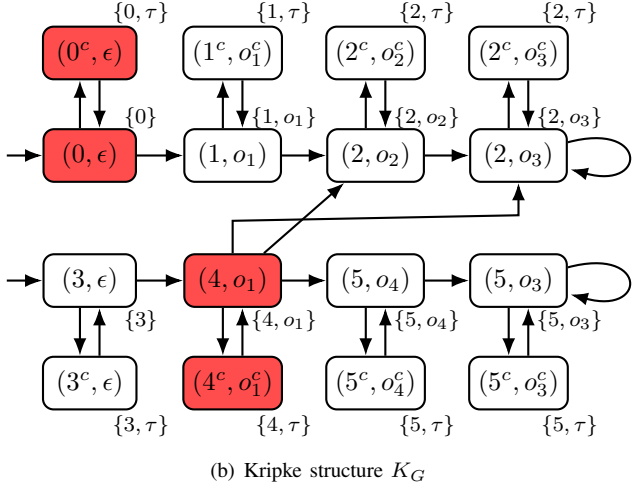 
	
We show the verification of initial-state opacity and current-state opacity together using the following example. 
\begin{myexm}\upshape
Let us consider system $G$ shown in Figure \ref{fig-exam-opacity-G}, where $X=\{0,1,2,3,4,5\}$, $X_\S\!=\!\{0,4\}$, $\Sigma\!=\!\{a,b,c,d,e\}$, $\O\!=\!\{o_1,o_2,o_3,o_4\}$, and the observation mask $M\!:\!\Sigma\!\to\!\O$ is defined by: $M(a)\!=\!M(b)\!=\!o_1,M(c)\!=\!o_2,M(d)\!=\!o_3,M(e)\!=\!o_4$. The modified Kripke structure $\tilde{K}_G$ is shown in Figure \ref{fig-exam-opacity-KG}. 
Clearly, system $G$ is initial-state opaque.  
To see how this is captured by $K_G$ (by omitting copies states in $\tilde{K}_G$) and $\phi_{iso}$, let us consider
\[
\pi_1\!=\!\{0\}\{1,o_1\}\{2,o_2\}\{2,o_3\}^\omega \in \trace(K_G),
\]
which is  the only trace such that $\S^\pi$ holds. Then we have  
\[
\pi_2\!=\!\{3\}\{4,o_1\}\{2,o_2\}\{2,o_3\}^\omega \in \trace(K_G)
\]
such that $\NS^\pi$ and $\square(o^{\pi_1}\!=\!o^{\pi_2})$ hold. 
Therefore $K_G\models \phi_{iso}$. 

Also, one can use Theorem~\ref{thm-cso} to check that system $G$ is also current-state opaque. 
For example, let us consider   trace 
$\pi_1\!=\!\{3\}\{4,o_1\}\{4,\tau\}\cdots\!\in\! \trace(\tilde{K}_G)$, 
where propositions after $\{4^c,\tau\}$ are omitted since they are irrelevant in Equation~\eqref{eq:cso-thm} as long as they do not contain $\tau$.  
We have that both $\lozenge_1 \tau^{\pi_1}$ and $\square (\tau^{\pi_1}\!\to\! \S^{\pi_1} ) $ hold for $\pi_1$. 
For such $\pi_1$, we can find 
trace 
$\pi_2\!=\!\{0\}\{1,o_1\}\{1,\tau\}\cdots\in \trace(\tilde{K}_G)$
such that 
(i) $(o^{\pi_1}\!=\!o^{\pi_2})\U \tau^{\pi_1}$ since they have the same observation before the occurrence of $\tau$; and 
(ii) 
$\square(\tau^{\pi_1}\!\to\! (\tau^{\pi_2}\!\wedge\! \NS^{\pi_2}) )$ 
since $\{1,\tau\}$ holds for $\pi_2$ when $\tau$ holds for $\pi_1$. 
Similarly, for any   trace $\pi_1\!=\!\{0\}\{0,\tau\} \cdots\in \trace(\tilde{K}_G)$, we can find trace $\pi_2\!=\!\{3\}\{3,\tau\}\cdots\in \trace(K_G)$ such that the condition in Equation~\eqref{eq:cso-thm}  holds. Therefore, we also have $\tilde{K}_G\models \phi_{iso}$. 
\end{myexm}

\subsection{Infinite-Step Opacity in HyperLTL}
The case of infinite-step opacity is similar to the case of current-state opacity. 
Specifically, it also requires that, 
for any string ending up with a secret state, there exists another string ending up with a non-secret state such that they have the same observation. 
However, in addition, we need to further ensure that, for any string starting from the above secret, 
there exists another string starting from the above non-secret state such that they have the same observation.  Otherwise, the intruder may realize that the system was at a secret state after some steps. 
To capture this difference, one can simply replace the 
``truncated" observation equivalence condition $ (o^{\pi_1}\!=\!o^{\pi_2})\U \tau^{\pi_1}$ in Equation~\eqref{eq:cso-thm} 
by an infinite horizon version of observation equivalence condition
$\square (o^{\pi_1}\!=\!o^{\pi_2})$.  
This leads to the following theorem.  
	\begin{tcolorbox} 
        \begin{mythm}[\textbf{HyperLTL for Infinite-Step Opacity}]\label{thm-ifo}\upshape  
        System $G$ is infinite-step opaque if and only if $\tilde{K}_G\models \phi_{ifo}$, 
        where
        \begin{align}
            &\phi_{ifo}= \\
            &\forall\pi_1.\exists\pi_2.
        \left[
        \begin{aligned}
            & \left[  
        \lozenge_1 \tau^{\pi_1} \wedge\square (\tau^{\pi_1}\!\to\! \S^{\pi_1} ) 
        \right]\to\\
            &[\square(o^{\pi_1}=o^{\pi_2})\wedge
            \square(\tau^{\pi_1}\to (\tau^{\pi_2}\wedge\NS^{\pi_2}))]\notag
        \end{aligned}
        \right].
        \end{align}
        \end{mythm}
    \end{tcolorbox} 

    \begin{proof}
    ($\Rightarrow$)
    Suppose, for contraposition, that $\tilde{K}_G\not\models\phi_{ifo}$, i.e., there exists a run
    \begin{flalign}
    \rho&=(x_0,\epsilon)(x_1,o_1)\cdots(x_k,o_k)(x^c_k,o^c_k)(x_k,o_k)\notag\\
    &~~~~(x_{k+1},o_{k+1})(x_{k+2},o_{k+2})\cdots\in\run(\tilde{K}_G)\notag
    \end{flalign}
    such that $\tilde{L}((x^c_k,o^c_k))\!=\!\{x_k,\tau\}$ and $x_k\!\in\! X_\S$, i.e., $\lozenge_1
    \tau^{\pi}\wedge\square(\tau^{\pi}\to\S^{\pi})$ holds for trace $\pi\!=\!\tilde{L}(\rho)$, 
    and for any other run
    \begin{flalign}
    \rho'&=(x'_0,\epsilon)(x'_1,o_1)\cdots(x'_k,o_k)(x'^c_k,o^c_k)(x'_k,o_k)\notag\\
    &~~~~(x'_{k+1},o_{k+1})(x'_{k+2},o_{k+2})\cdots\in\run(\tilde{K}_G)\notag
    \end{flalign}
    i.e., $\square(o^{\pi}\!=\!o^{\pi'})\wedge\square(\tau^{\pi}\!\to\!\tau^{\pi'})$ holds for trace $\pi'\!=\!\tilde{L}(\rho')$, we have  $x'_k\!\in\! X_\S$.
    Now, for each $j\!\geq\! k$, we define $\hat{q}_j^k$ as the delayed-state estimate of $o_1\cdots o_j$ for observation $o_1\cdots o_k$, i.e.,
    \begin{equation}\label{eq-qjkhat}
    \hat{q}_j^k=
    \left\{ x'_k\in X
    : \!\!\!\!\!\!\!
    \begin{array}{cc}
        &  (x'_0,\epsilon)(x'_1,o'_1)\cdots\in \run(\tilde{K}_G)\\
        & 
    \text{ s.t. }
    o'_i=o_i,\forall 1\leq i\leq j
    \end{array}
    \right\} 
    \end{equation}
    By construction, we have:
     (i) $\hat{q}_j^k\!=\!\hat{X}_G(o_1\cdots o_k\mid o_1\cdots o_j)$; and
     (ii) $\hat{q}_{j+1}^k\!\subseteq\!\hat{q}_j^k$.
    For infinite sequence $\hat{q}_k^k\hat{q}_{k+1}^k\cdots\!\in\!(2^X)^\omega$, we define
    \[
    \X_{\inf}^k=\{\hat{q}^k:\forall n\in\NN,\exists j>n\text{ s.t. }\hat{q}_j^k=\hat{q}^k\}
    \]
    as the set of delay-state estimates that appear infinite number of times in $\hat{q}_k^k\hat{q}_{k+1}^k\cdots$. Let $n_{\inf}$ be the smallest integer such that $\hat{q}_{n_{\inf}}^k\hat{q}_{n_{\inf+1}}^k\cdots\in(\X_{\inf}^k)^\omega$. We claim: for each $\hat{q}^k\in\X_{\inf}^k$, we have $\hat{q}^k\subseteq X_\S$. Otherwise, there would exist such a run $\rho'$ that $o'_i=o_i,\forall i\geq1$ but $x'_k\notin X_\S$.
    Now, let us consider the observations $\alpha\!=\!o_1\ldots o_k$ and $\beta\!=\!\alpha o_{k+1}\ldots o_{n_{\inf}}$. It holds that $\hat{X}_G(\alpha\mid\beta)\!=\!\hat{q}_{n_{\inf}}^k\subseteq X_\S$, which makes $G$ not infinite-step opaque.
    
    ($\Leftarrow$)
    We still prove this direction by contraposition. Suppose that $G$ is not infinite-step opaque, which means that there exist observations $\alpha\!=:\!o_1\cdots o_k,\beta\!=:\!\alpha o_{k+1}\cdots o_n\in M(\L(G))$ such that $\hat{X}_G(\alpha\mid\beta)\subseteq X_\S$. Since $G$ is live, one can easily obtain an infinite string $\beta'=\beta o_{n+1}\cdots\in M(\L^\omega(G))$. 
    Based on construction of $\tilde{K}_G$, there exists a run
    \begin{flalign}
    \rho&=(x_0,\epsilon)(x_1,o_1)\cdots(x_k,o_k)(x^c_k,o^c_k)(x_k,o_k)\notag\\
    &~~~~(x_{k+1},o_{k+1})\cdots(x_n,o_n)\cdots\in\run(\tilde{K}_G)\notag
    \end{flalign}
    where $\tilde{L}((x^c_k,o^c_k))\!=\!\{x_k,\tau\}$ and $x_k\!\in\! X_\S$, i.e., 
    $\lozenge_1\tau^{\pi}\wedge\square(\tau^{\pi}\to\S^{\pi})$ holds for trace $\pi\!=\!\tilde{L}(\rho)$.
    Furthermore, we have $\hat{q}_n^k\!=\!\hat{X}_G(\alpha\mid\beta)\!\subseteq\! X_\S$, where $\hat{q}_n^k$ is defined in Equation \eqref{eq-qjkhat}.
    Then it holds that $\hat{q}_j^k\subseteq\hat{q}_n^k\subseteq X_\S$ for all $j\geq n$. Now, we claim that the following holds
    \begin{equation}\label{eq-notifo}
        \forall\pi'\in\trace(\tilde{K}_G).\left[
        \begin{aligned}
            &[\square(o^{\pi}=o^{\pi'})\wedge\square(\tau^{\pi}\to\tau^{\pi'})]\\
            &~~\to[\square(\tau^{\pi}\to(\tau^{\pi'}\wedge\S^{\pi'}))]
        \end{aligned}
        \right]
    \end{equation}
    Otherwise, it means that there would be a run 
    \begin{flalign}
    \rho'&=(x'_0,\epsilon)(x'_1,o_1)\cdots(x'_k,o_k)(x'^c_k,o^c_k)(x'_k,o_k)\notag\\
    &~~~~(x'_{k+1},o_{k+1})\cdots(x'_n,o_n)\cdots\in\run(\tilde{K}_G)\notag
    \end{flalign}
    where $x'_k\in X_\NS$. Then we have $\{x_k,x'_k\}\subseteq\hat{q}_j^k\not\subseteq X_\S$ for all $j\geq k$. This immediately contradicts with $\hat{q}_j^k\subseteq X_\S,\forall j\geq k$. Therefore, we know that Equation \eqref{eq-notifo} holds, which shows that $\tilde{K}_G\not\models\phi_{ifo}$. The proof is thus completed.
    \end{proof}

\begin{myexm}\upshape
Still, let us consider   system $G$  in Figure \ref{fig-exam-opacity-G}, whose modified Kripke structure $\tilde{K}_G$ is shown in Figure \ref{fig-exam-opacity-KG}. 
Let us consider the following trace
\[
\pi_1=\{3\}\{4,o_1\}\{4,\tau\}\{4,o_1\}\{5,o_4\}(\{5,o_3\})^\omega \in \trace(\tilde{K}_G)
\]
For $\pi_1$, we have $\lozenge_1 \tau^{\pi_1} \wedge\square (\tau^{\pi_1}\!\to\! \S^{\pi_1} ) $ 
since $4\in X_\S$ is a secret state. 
However, for any trace $\pi_2$ satisfying 
$\square(\tau^{\pi_1}\!\to\! (\tau^{\pi_2}\!\wedge\! \NS^{\pi_2}) )$, e.g., 
\[
\pi_2= \{0\}\{1,o_1\}\{1,\tau\}\{1,o_1\}\{2,o_2\}(\{2,o_3\})^\omega \in \trace(\tilde{K}_G),
\]
condition $\square (o^{\pi_1}\!=\!o^{\pi_2}) $ does not hold. 
Therefore, we have $\tilde{K}_G\not\models \phi_{ifo}$, i.e.,  $G$ is not infinite-step opaque by Theorem \ref{thm-ifo}.  
\end{myexm}

\section{Concluding Discussions}\label{sec-conclusion}
In this paper, we revisited the problems of verifying  observational properties for partially-observed DES, which have been studied very actively and extensively in the past two decades in the context of DES. We showed that the recent developed new temporal logic called HyperLTL can be used as a suitable tool for unifying many of the important observational properties in the literature. Our framework does not provide new decidability results since  for all properties considered here, verification algorithms have already been developed. However, we believe that our unified view in terms of HyperLTL provides new insights for those properties that were previously investigated separately. Furthermore, our unified framework is of practical value since it provides the access to many of the efficient model checking tools for the purpose of verifying all these observational properties instead of developing a customized algorithm for each case.  

Among the properties investigated in this paper, it has been shown before that deciding all notions of opacity as well as weak detectability are PSPACE-complete, while the remaining notions such as diagnosability and strong detectability can all be decided in polynomial-time. Now let us go back to the second question in the introduction that why some properties are similar while some are more different. This, in fact, can be easily explained by the theory of HyperLTL.  Particularly, for a HyperLTL formula, the \emph{alternation depth} is referred to as the number of times the quantifiers alter from existential to universal, or vice versa. It has been shown that the verification complexity of HyperLTL will increase an exponential level when the formula has one more alternation of quantifiers \cite{clarkson2014temporal}. For example, the alternation depth of formulae ``$\forall.\exists.$" is one, while the alternation depth of formulae  ``$\forall.\forall.$" is zero.  The former corresponds to the case of opacity and weak detectability, while the latter corresponds to other notions such as diagnosability. 
Therefore, expressing observational properties in terms of HyperLTL also suggests a natural way to classify existing notions of observational properties in partially-observed DES:  the larger alternation depth the property has, the higher verification complexity it will require.  

Finally, we would like to remark that, although we have shown that many of the important properties in partially-observed DES can be formulated in terms of HyperLTL, there still exist properties that cannot be captured by our framework. 
Example of such properties are A-diagnosability \cite{thorsley2005diagnosability,chen2018revised} and A-detectability \cite{keroglou2015detectability}. For example, A-diagnosability requires that after the occurrence of a fault, at each state, there always \emph{exists a path} along which the fault can be detected. Essentially, this property cannot be captured by HyperLTL, since  HyperLTL  still belongs to the category of linear-time properties (although it is evaluated over multiple traces). The existence of a path from a state satisfying some condition is essentially a \emph{branching-time property} which is beyond the semantics of HyperLTL. To address this issue and to further generalize our framework, a possible direction is to use more expressive temporal logic such as HyperCTL$^*$  \cite{clarkson2014temporal} that supports both linear-time and branching-time properties over multiple traces. We plan to investigate this direction in our further work. 

\bibliographystyle{plain}
\bibliography{mybib_abbrv}

\end{document}